\documentstyle[aps,preprint]{revtex}
\input amssym.def
\input amssym.tex
\everymath{\rm}
\everydisplay{\rm}
\begin{document}
\title
{On the Fermi Liquid to Polaron Crossover II: \\
Double Exchange and the Physics of ``Colossal'' Magnetoresistance}
\author{A. J. Millis, R. Mueller and Boris I. Shraiman}
\address{Bell Laboratories \\
Lucent Technologies\\
600 Mountain Avenue  \\
Murray Hill, NJ 07974 }
\maketitle
\renewcommand{\baselinestretch}{2}
\begin{abstract}
We use the dynamical mean field method to study
a model of electrons Jahn-Teller coupled to localized
classical oscillators and ferromagnetically
coupled to ``core spins'', which, 
we argue, contains the essential
physics of the ``colossal magnetoresistance'' manganites
$Re_{1-x} A_x MnO_3$.
We determine the different regimes of the model and
present results for the temperature
and frequency dependence of the conductivity, 
the electron spectral
function and the root mean square lattice parameter
fluctuations.  We compare our results
to data, and give a qualitative discussion of important
physics not included in the calculation.
Extensive use is made of results from a companion paper
titled:
``On the Fermi Liquid to Polaron Crossover I:  General Results''.
\end{abstract}
\newpage

\section{Introduction}
The doped rare earth manganites have been studied for
many years \cite{Wollan55,Goodenough55} and
interest in the materials has revived following
 the recent
discovery of extremely large magnetoresistance in some
members of the family \cite{Jin94}.
The chemical formula is $Re_{1-x}A_xMnO_3$, with Re a
rare earth element such as La or Nd and A a divalent metal
ion such as Sr or Ca.
The electronically active orbitals are believed to be
the Mn d-orbitals \cite{Wollan55,Goodenough55} and the
mean d-occupancy is $4-x$.
Each Mn ion feels an approximately cubic crystal field 
which splits the Mn
d-levels into a $t_{2g}$ triplet and an $e_g$ doublet
\cite{grouptheory}.  The $t_{2g}$ levels are believed
\cite{Goodenough55,Mattheiss94} to lie
substantially ($\sim$ 5eV) below the $e_g$ levels.
On-site Coulomb interactions are  apparently
strong enough  that no d-orbital may be occupied by more than
one electron.  Further, all electron spins
in Mn-$d$ orbitals are aligned by a 
large ferromagnetic Hunds-rule coupling.
The Coulomb and Hunds-rule 
interaction energies have not been measured
directly, but there is substantial indirect evidence that
they are large.
For example, at $0.2 \lesssim x \lesssim 0.5$ (precise
values depend on Re and A) the ground state is ferromagnetic,
and the observed magnetization is consistent with all
$4-x$ electrons on each Mn being lined up in the maximal
spin state \cite{Wollan55}, suggesting a large Hunds coupling.
Also, $ReMnO_3$ undergoes a structural phase
transition at $T \approx 800$K, which has been shown
\cite{Kanamori61,Ellemans71} to be due to a staggered
($\pi , \pi , \pi$) ordering of Jahn-Teller distortion
of locally $e_g$ symmetry.
This would not occur unless the $e_g$ orbital were 
singly occupied, which in turn implies that the $t_{2g}$
orbitals are also singly occupied, suggesting a large 
on-site Coulomb interaction.

The resulting physical picture 
is that $3$ of the $(4-x)$ d electrons fill up the
$t_{2g}$ levels, forming an electrically inert core
spin $\vec{S}_c$ of magnitude $S_c=3/2$.
The remaining $1-x$ electron goes into a
linear combination of the $e_g$ levels, and is coupled
to $\vec{S}_c$ by a Hunds rule coupling $J_H$ which is presumably
large, but has not been directly measured.
Okimoto et al. have recently presented an interpretation
of optical data implying that $J_HS_c\sim 1.2$eV
\cite{Okimoto95}.
We shall argue below that their
interpretation is not correct and that $J_HS_c$ is rather larger.
Certainly, the conventional \cite{Wollan55}
wisdom is that the limit
$J_HS_c \rightarrow \infty$ is appropriate, so one only need
consider configurations with $e_g$ electrons parallel to
core spins.

The $Re_{1-x}A_xMnO_3$ materials display
a wide range of interesting physics.
For $0 \leq x \lesssim 0.2$ (all x-values are
approximate, and depend on Re and A) the materials are
insulating at all temperatures and are antiferromagnetic
or ferrimagnetic at low T.
For $0.2 \leq x \leq 0.5$ the low-T
phase is a fully polarized ferromagnetic metal.
As the temperature is increased for 
$0.2 \leq x \leq 0.5$, there is a ferromagnet
to paramagnet transition, which may be of first or
second order, at a $T_c(x) \sim 300$K.
In the paramagnetic state the material may be
either ``metallic'' (in the sense that
$d \rho /dT > 0$ and $\rho \lesssim \rho_{Mott}$)
or ``insulating'' (in the
sense that $d \rho /dT <0$ and
$\rho \gtrsim \rho_{Mott}$).  (Here
$\rho_{Mott}$, the Mott ``maximum metallic resistivity'',
is about $\sim  1000 \mu \Omega$-cm and
corresponds to a mean free path of order $p_F^{-1}$
\cite{Kastner87}).
Insulating behavior occurs at lower x and metallic
behavior at higher x.
There is for all x a very pronounced drop in
$\rho$ as T is lowered through $T_c$, and in this
regime the resistivity has a very strong magnetic
field dependence.
The ``colossal'' magnetoresistance of interest
here occurs 
for $x$ such that 
the material is in the ``insulating'' regime at
$T > T_c$ but is  a metallic ferromagnet at
$T < T_c$.
Finally, at $x \gtrsim 0.5$ the low T state
is charge-ordered, antiferromagnetic and insulating
\cite{Wollan55,Chen94}.  We do not address the
physics of this regime here.
A qualitative phase diagram is shown in Fig. 1.

Most models \cite{Anderson54,deGennes61,Searle71,Kubo83,Furukawa94}
of electron transport in $Re_{1-x}A_xMnO_3$  have emphasized
the ``double exchange'' phenomena caused by the large
Hunds coupling $J_H$.
The essence of double exchange is that 
when an electron hops from site $i$ to site $j$
it must also go from having its spin parallel to
$\vec{S}_c^i$ to having its spin parallel to
$\vec{S}_c^j$;  the hopping amplitude, $t_{ij}$, thus depends
upon relative spin orientation \cite{Anderson54}.
For two fixed sites $i$ and $j$ it is possible to
choose phase factors so that
$t_{ij} \rightarrow \frac{t_{ij}}{\sqrt{2}}
\sqrt{1+ \frac{S_{c}^i \cdot S_{c}^j}{S_c^2}} \equiv t_{ij}
\cos [ \theta_{ij}/2]$.

The double exchange phenomenon gives an obvious
connection between electron hopping and magnetic
order:  disorder in the spins implies 
randomness in $t_{ij}$, which decreases 
below $T_c$ or in a field. 
This effect seems very likely to be related to the
``colossal'' magnetoresistance observed near $T_c$.
However, two of us  and P. B. Littlewood have 
recently argued that models involving
only double exchange cannot explain the observed resistivity
\cite{Millis95a}.
The essential point is that in materials
exhibiting ``colossal'' magnetoresistance the
resistivity at $T >T_c$ is much larger than the
Mott limit and moreover rapidly increases as T decreases.
Indeed, as shown in Appendix A, the observed $T>T_c$ resistivities
are so large that a classical description involving
particles incoherently hopping from site to site with
a hopping probability $W \ll k_BT/ \hbar$ is appropriate.
In models involving only double exchange the scattering
produced by spin disorder is simply not large enough
to cause such insulating behavior.
A straightforward calculation
\cite{Kubo83,Millis95a} shows that if the spins are
completely decorrelated one finds
$p_F\ell \sim 3$,
i.e. $W \sim t_{ij} \gg kT$.
More sophisticated arguments involving localization and
phase factors are shown in Appendix B not to change this
conclusion significantly.
Therefore we believe some additional physics not included in the
double exchange-only model must be important.
This conclusion is not universally 
accepted \cite{Furukawa94,Varma96}

One possible source of this extra physics is the
``Hubbard U'' effect of the on-site Coulomb interaction
which produced the Hunds coupling in the first place.
While this is undoubtedly quantitatively important,
we do not believe it is the primary cause of the observed
insulating behavior, essentially because away from commensurate
densities (such as one electron per site) 
canonical Mott insulating materials such as the high
$T_c$ superconductors or other doped transition-metal oxides
have resistivities which are rather less than the
Mott limit and which decrease with temperature
\cite{Hwang92,Tokura93}, in stark contrast to the
behavior observed at $ T > T_c$ in $Re_{1-x}A_xMnO_3$.

We proposed \cite{Millis95a} that the crucial additional
physics is a strong electron-phonon coupling, which localizes
the conduction electrons as polarons at $T > T_c$ and smaller $x$,
but  is weakened
in the $T < T_c$ ferromagnetic state,
restoring metallic behavior.
We argued that this is possible because the
behavior of electron-phonon model is controlled by a
dimensionless coupling parameter which is the ratio
of an interaction energy to the electron kinetic
energy.
The double exchange physics implies that ferromagnetic
order increases the electron kinetic energy,
thereby decreasing the effective coupling strength.
Also, a recent analysis \cite{Millis96a} 
of the structural distortion
observed \cite{Ellemans71} in $LaMnO_3$ showed that the
electron-phonon coupling is indeed strong.

In this paper we present a detailed study of a model of
electrons coupled to core spins and to phonons which
we believe confirms the importance of electron-phonon
interactions.  We use a "dynamical mean field"
method which has previously been extensively applied to 
interacting problems without double exchange \cite{Georges96} 
and has been used
by Furukawa to study models involving only double
exchange \cite{Furukawa94}.
The model we study does not capture all of the
physics of $Re_{1-x}A_xMnO_3$; in particular, Coulomb
effects 
and quantum and intersite terms in the
phonon Hamiltonian are omitted and an oversimplified electron-phonon
coupling is used.
We therefore cannot quantitatively compare our results
to experiment.
The qualitative agreement we obtain is however
compelling.

Other workers
have also studied electron-phonon
effects in manganites and related materials.  
Emin, Hillery and Liu
studied a theoretical model of a single bound polaron
coupled to spin waves and found a temperature dependence of
the polaron size which they argued could be related to
transport anomalies at $T_c$  observed
in EuO \cite{Emin86}.  Their work, however, is based
on models in which double exchange physics
is not relevant, reflecting the different physics of EuO.
Roder, Zang and Bishop used 
variational wavefunction
techniques to examine the interplay between electron-phonon
interaction
and  double exchange \cite{Roder96}.  Their 
work is in a sense complementary
to ours.  They have incorporated quantum phonons
and have presented some results on intersite phonon correlations,
but their technique seems to work best at low temperatures.
To calculate properties at and above $T_c$ they resort
to a dilute limit approximation which amounts to
the study of a single carrier in a deformable medium.
Also,  they have not presented
results for transport and optical quantities; their main result is a
calculation of the coupling dependence and doping dependence
of $T_c$.  Their results for the coupling dependence are very similar
to ours; their results for the doping dependence are
based on an assumption which, we argue, is not justified
by the arguments in their paper
A comparison of their results for $T_c$
to ours is given in Section IV,
and a further discussion is given in the conclusion.

The rest of this paper is organized as
follows.
In Section II we define the model and the approximations
used.
In Section III we give the qualitative physics of the
model, distinguishing different regimes and
presenting the behavior of physical quantities in each.
In Section IV we present the
results of a detailed numerical study of the model
at half-filling.
In Section V we present and discuss results at other dopings.
Section VI is a conclusion, in which the relation of
results to data is analysed and the effects of omitted
interactions are outlined.
Appendix A discusses the theoretical interpretation of the observed
resistivities, in particular the relation to the Mott minimum
metallic conductivity and to transport by classical particles
while Appendix B discusses in more detail the resistivity
of the double exchange-only model.
An announcement of some of the results
of this paper has been submitted
elsewhere \cite{Millis95b}.

This paper relies heavily on results of a 
companion paper which uses the dynamical mean field method
to study
electron-phonon interactions
in models without double exchange
\cite{Millis96b}, to which we refer henceforth as I.

\section{Model and Approximations}
We study a model of Mn $e_g$ d-electrons coupled to core spins
$\vec{S}_c^i$ and phonons, with Hamiltonian
\begin{equation}
H = H_{band} + H_{d-ex} + H_{el-ph} + H_{ph}.
\label{eq:H}
\end{equation}
$H_{band}$ describes the hopping
of d-electrons  between
sites $i,j$ of a lattice.  We take
the electrons to have two-fold orbital degeneracy labelled
by a Roman index (a,b) and two fold spin degeneracy
labelled by a Greek index $( \alpha , \beta )$.  Explicitly,
\begin{equation}
H_{band}= - \sum_{\langle ij \rangle ab\alpha} t_{ij}^{ab}
d_{ia\alpha}^{\dagger} d_{jb\alpha}
- \mu \sum_{ia\alpha}
d_{ia\alpha}^{\dagger} d_{ia\alpha}.
\label{eq:Hband}
\end{equation}
The hopping matrix element $t_{ij}^{ab}$ is a real
symmetric matrix whose form depends on the choice
of basis in $ab$ space and the direction of the $i-j$ bond.
The precise form will not be important in what follows.

The interaction of itinerant electrons
with the core spins is given by
\begin{equation}
H_{d-ex} = -J_H\sum_{ia\alpha\beta} \vec{S}_c^i\cdot
d_{ia\alpha}^{\dagger}\vec{\sigma}_{\alpha\beta}
d_{ia\beta}
\label{eq:H_{dex}}
\end{equation}
As discussed below, we shall take the limit
$J_HS_c\rightarrow\infty$.

We assume a Jahn-Teller form for the electron-phonon coupling.
Previous analysis  \cite{Kanamori61,Millis96a}
has shown that this coupling is strong in $LaMnO_3$,
so it may be expected to be strong also in doped
compounds.
Thus we focus on lattice distortions which split the
on-site orbital degeneracy of the $e_g$ levels.
Physically, these correspond to $e_g$ symmetry distortions of
the oxygen octahedra around an Mn site.
Mathematically, we may parametrize a local $e_g$ distortion
by a magnitude $r$ and an angle $\theta$,
and define a two-component vector
$\vec{r}=(r_z,r_x)$ with $r_z=r\cos\phi$ and
$r_x =r\sin\phi$.
The coupling of this to the $e_g$ levels is prescribed
by group theory \cite{grouptheory} to be
\begin{equation}
H_{el-ph} = g \sum_{iab\alpha} \vec{r}_i \cdot
d_{ia\alpha}^+ \vec{\tau}^{ab} d_{ib\alpha}.
\label{eq:Helph}
\end{equation}
Here $\vec{\tau}=(\tau^z,\tau^x)$ is a vector of Pauli
matrices acting in orbital space.

It is important to note that the coupling written in
Eq. \ref{eq:Helph} is not the only physically relevant one.
In ref \cite{Millis96a} it was argued that a Mn site
with no $e_g$ electrons would induce a breathing distortion
of the surrounding oxygen ions, and that this breathing
distortion played an important role in determining the x dependence
of the structural phase boundary.
We have not included this coupling in the present
calculations, but will qualitatively discuss its effects in
the conclusion.

In order to obtain a tractable model we assume $H_{ph}$ describes
classical harmonic oscillators of stiffness $k$ which are
furthermore independent from site to site.  Thus,
\begin{equation}
H_{ph}=\sum_i \frac{1}{2} kr_i^2.
\label{eq:Hph}
\end{equation}

Despite the simplifying approximations, the model
defined by Eq. \ref{eq:H} is not solvable except in certain
limits.
To obtain results, we adopt the ``dynamical mean field''
approximation, which becomes formally exact in a limit
in which the spatial dimensionality $d\rightarrow\infty$ and has
been shown to provide a reasonable description of models of
interacting electrons in $d=3$ \cite{Georges96}.
Recently, the technique has also been applied to the
double exchange-only model defined by Eq. \ref{eq:H} with
$H_{el-ph}=0$ \cite{Furukawa94}.

The dynamical mean field method is based on an assumption
about the electron Green function $G_{\alpha\beta}^{ab}(p,\omega)$.
In general this is a tensor in spin and orbital space 
which may be written
\begin{equation}
\left[ G_{\alpha\beta}^{ab} (p,\omega) \right]^{-1} =
\omega-\epsilon_p^{ab}+mu \Sigma_{\alpha\beta}^{ab}
(p,\omega)
\label{eq:G}
\end{equation}
Here $\epsilon_p^{ab}$ is the dispersion implied by
Eq. \ref{eq:Hband} and $\Sigma$ is the self energy
due in the present problem to $H_{el-ph}$ and $H_{d-ex}$.
The fundamental approximation of the dynamical mean
field method is the neglect of the momentum (p) dependence
of $\Sigma$.
This is a reasonable approximation because models of the
form of Eq. \ref{eq:H} (such as the usual
Migdal-Eliashberg electron-phonon Hamiltonian)
generally lead to a self energy with a weak momentum
dependence in $d=3$ \cite{Abrikosov63}.
If the momentum dependence of $\Sigma$ may be neglected,
then all physical quantities may be
expressed as functionals of the momentum-integrated Green
function $G_{LOC}$, given by
\begin{equation}
G_{LOC \alpha \beta}^{ab} ( \omega ) =
\int \frac{d^3p}{(2 \pi )^3}
G_{\alpha \beta}^{ab} (p, \omega )
\label{eq:GLOCab}
\end{equation}
We shall assume that there is no long range order in
orbital space, so $G_{LOC}$ and also
$\Sigma (\omega )$ must be proportional to the unit
matrix in $ab$ space.
We shall allow for the possibility of ferromagnetic order,
and shall take the ordered moment parallel to $z$.
We may then write
\begin{equation}
G_{LOC \alpha \beta}^{ab}(\omega )=g_0(\omega ) +
g_1(\omega )\sigma^z
\label{eq:GLOC}
\end{equation}
and a similar equation for $\Sigma$.
We simplify Eq. \ref{eq:GLOCab} by first writing
$g_{0,1}= \frac{1}{4}Tr_{ab}Tr_{\alpha \beta}
\int \frac{d^3p}{(2\pi )^3}G_{\alpha \gamma}^{ab}
(\sigma_{\gamma \beta}^z)^{0,1}$, then 
introducing at
each $p$ the rotation matrix
$R_p^{ab}$ which
diagonalizes $\epsilon_p^{ab}$, i.e.
\begin{equation}
\epsilon_p^{ab}=R_p
\left[
\begin{array}{cc}
\epsilon_p^1 & 0 \\
0 & \epsilon_p^2
\end{array}
\right]
R_p^{-1}
\label{eq:diag}
\end{equation}
and finally exploiting the cyclic invariance of the trace.
We obtain
\begin{equation}
\begin{array}{rl}
g_0(\omega) &= \frac{1}{4} Tr_{ab} Tr_{\alpha\beta} \int
d\epsilon_p {\cal D} (\epsilon_p) [\omega -\epsilon_p +\mu-
\Sigma_{\alpha \beta}( \omega )]^{-1} \\
g_1(\omega ) &= \frac{1}{4} Tr_{ab} Tr_{\alpha \beta} \int
d\epsilon_p {\cal D} (\epsilon_p) \sigma^z 
[\omega - \epsilon_p +\mu-
\Sigma_{\alpha \beta} ( \omega )]^{-1}
\label{eq:g}
\end{array}
\end{equation}
where $\epsilon_p$ is either of the eigenvalues
of $\epsilon_p^{ab}$ (they are related by symmetry operations)
and ${\cal D}$ is the density of states, which we take to be
semicircular with width $D=4t$:
\begin{equation}
{\cal D} (\epsilon_p) = 
\sqrt{4t^2-\epsilon_p^2}  /2\pi t^2.
\label{eq:D}
\end{equation}
Because $G_{LOC}$ is momentum independent and involves two
independent functions, it must be the Green function of some
effective single site model involving two mean field functions
$a_0$ and $a_1$.
This model is described by the partition function
\begin{equation}
Z_{LOC} = \int rdrd \phi \int d^2
\vec{\Omega}_c exp [S_{LOC}]
\label{eq:ZLOC}
\end{equation}

Here $r$ and $\phi$ are the classical
oscillator coordinates introduced above
Eq. \ref{eq:Helph},
$\vec{\Omega}_c = \vec{S}_c / |S_c|$, and the
integrals are simple integrals rather than functional
integrals because we have taken $r$,$\phi$
and $\vec{\Omega}_c$ to
be classical.

The effective action $S_{LOC}$ is
\begin{equation}
\begin{array}{rl}
S_{LOC} =- \frac{1}{2} \frac{k}{T} r^2 &+
Tr ln [a_0(i\omega_n)+a_1(i\omega_n)
\sigma_z + J_H \vec{\Omega}_c \cdot
\vec{\sigma}_{\alpha\beta} \\
   &+ g \vec{r} \cdot \vec{\tau}_{ab}] - \vec{h}_{ext} \cdot
S_c \vec{\Omega}_c /T.
\label{eq:SLOC}
\end{array}
\end{equation}

Here we have added a term coupling the core
spin to an external field $\vec{h}_{ext}$.
One could also couple the external field to the $e_g$
electrons, but the factor of $3/(1-x)$ in size of moment
means that this coupling is unimportant.

The mean field parameters $a_0$, $a_1$ 
in Eq. \ref{eq:SLOC} are determined
 as follows \cite{Georges96}.
One obtains the local Green functions
$g_{0,1}^{LOC} =\frac{1} {4} \delta ln[Z_{LOC}] / \delta a_{0,1}$,
defines from these self energies
$\Sigma_{0,1} = a_{0,1} - ( g_{0,1}^{LOC})^{-1}$,
and demands that
$\Sigma_{\alpha\beta} \equiv \Sigma_0 + \Sigma_1 \sigma_z$,
 reproduces $g_{0,1}^{LOC}$ when used in Eqs. \ref{eq:g} and
\ref{eq:GLOC}.
For the semicircular density of states the resulting
equations may be written
\begin{equation}
\begin{array}{rl}
a_0 (\omega) &= \omega + \mu - \frac{t^2}{4}
\frac{\delta ln[Z_{LOC}]}{\delta a_0(\omega)}, \\
a_1(\omega) &= - \frac{t^2}{4}
\frac{\delta ln[Z_{LOC}]}{\delta a_1(\omega)}.
\label{eq:MF}
\end{array}
\end{equation}
The factor of four is
that appearing in Eq. \ref{eq:g}.

These equations simplify in the ``double exchange'' limit
$J_HS_c \rightarrow \infty$.
The argument of the $Tr$ $ln$ 
in Eq. \ref{eq:SLOC} is a matrix in the direct
product of spin and orbital space.
It has four eigenvalues, $a_0 \pm \Delta \pm gr$ with
$\Delta =|a_1 \hat{z} + J_HS_c \vec{\Omega}|$.
These are, of course, independent of the angle
$\phi$ describing the phonon.
For $J_HS_c \gg t$ the eigenvalues at
$a_0- \Delta \pm gr$ correspond to high energy states
which do not affect low energy phenomena.
Further, from Eq. \ref{eq:MF} it is clear that $a_1$ is of order
$t$, so $a_1 \ll J_HS_c$ and we may approximate
$\Delta \approx J_HS_c +a_1 \Omega_z$.
We may then absorb the constant term $J_HS_c$ into $a_0$
and $\mu$, rescale the parameters and define the spin angle
$\theta$ via $\Omega_z=cos(\theta)$ obtaining
\begin{equation}
S(x,\theta) =- \frac{x^2}{2T} + \sum_n ln [
(b_0+b_1 cos \theta)^2 - \lambda x^2 ]
+ h_0 cos \theta /T .
\label{eq:hS}
\end{equation}

Here $x = r \sqrt{k/t} $,
$b_{0,1} =a_{0,1} /t$, $\lambda =g^2 /kt$,
$h_0 = h_{ext} S_c /t$, and
$T$, $\omega$ and $\mu$ are measured
in units of $t$.
The mean field equations become
\begin{equation}
\begin{array}{rl}
b_0 &= \omega + \mu - \frac{1}{2}\int_0^{\infty} xdx 
\int_{-1}^{1} d(cos \theta ) P(x,\theta)
\frac{(b_0+b_1 cos\theta)}{(b_0+b_1\cos\theta)^2-\lambda x^2} \\
b_1 &= -\frac{1}{2}
\int_0^{\infty} xdx
\int_{-1}^{1} d( cos \theta )
P(x, \theta ) cos \theta
\frac{(b_0+b_1 cos \theta)}{(b_0+b_1 cos \theta)^2-\lambda x^2}
\label{eq:MF2}
\end{array}
\end{equation}
with
\begin{equation}
P(x, \theta ) = \frac{1}{Z_{LOC}} exp[S_{LOC}(x,\theta)]
\label{eq:P}
\end{equation}

These equations differ from those discussed in I by the presence
of the angular integral and by the quantity $b_1$, which
expresses the spin dependence of G.
Expressions for physical quantities are also slightly
different from those used in I because we must keep track
of the spin dependence.

The momentum integrated Green functions has components parallel
$( \uparrow \uparrow )$ and
antiparallel $(\downarrow \downarrow)$ to
the magnetization.
The off-diagonal
$(\uparrow \downarrow )$ components vanish.
We have
\begin{eqnarray}
G_{LOC}^{\uparrow \uparrow} ( \omega ) &=& 
\omega + \mu -b_0( \omega ) -b_1 ( \omega ) \\
G_{LOC}^{\downarrow \downarrow} ( \omega ) &=&
 \omega + \mu -b_0( \omega ) +b_1 ( \omega )
\label{eq:GLOCUP}
\end{eqnarray}
We shall be interested in the spectral function
\begin{equation}
A( \omega )=-Tr Im {\bf G}_{LOC} ( \omega+i\delta ) / \pi
\label{eq:A}
\end{equation}
The conductivity is given by
\begin{equation}
\sigma(i\Omega ) = \frac{2}{i\Omega} \int d \epsilon_p
{\cal D} (\epsilon_p) T \sum_{i\omega} Tr
[ {\bf G} (p, i \omega ) {\bf G} (p,i\omega +i\Omega )]
\label{eq:sigma}
\end{equation}
where the factor of two comes from the trace over
orbitals.  Here ${\bf G}$ is a diagonal matrix in spin space
and we have set $e=t=1$.

Another interesting quantity is the electron kinetic energy,
$K$, defined by
\begin{equation}
K = Tr_{ab} Tr_{\alpha\beta} \int \frac{d^3p}{(2\pi)^d}
\epsilon_p^{ab} \langle d_{pa\alpha}^+ d_{pb\beta} \rangle.
\label{eq:Kdef}
\end{equation}

By use of the relation between the expectation value and the
electron Green function, of Eq. \ref{eq:G} with
momentum independent self energy, and of the arguments
leading from Eq. \ref{eq:GLOCab} to
Eq. \ref{eq:g} and the mean field equations, we obtain
\begin{equation}
K = 2T \sum_n \left[ G_{LOC}^{\uparrow \uparrow} (\omega_n) \right]^2
+ \left[ G_{LOC}^{\downarrow \downarrow} (\omega_n) \right]^2.
\label{eq:K}
\end{equation}

The magnetization $m$ is given by
\begin{equation}
m = \int_0^{\infty} xdx \int_{-1}^{1} d \cos \theta \cos \theta
P(x, \theta ).
\label{eq:M}
\end{equation}
In these units the $T=0$ value of $m=1$.

We shall also be interested in the mean-square lattice
distortion, $\bar{x}^2$, given by
\begin{equation}
\bar{x}^2 = \int_0^{\infty} xdx \int_{-1}^{1} 
d( cos \theta) x^2 P(x, \theta ).
\label{eq:xbar}
\end{equation}

We conclude this section by mentioning numerical methods.
We use the procedures described in I, and handle the
additional angular integral by a twenty-point Legendre formula.
Computations are of course more time consuming because of
the extra integral involved.
We found it convenient  first to locate the magnetic
transition temperature $T_c$ and then to perform calculations
at $T>T_c$ using equations obtained by forcing $b_1=0$.
Convergence difficulties arise for temperatures near $T_c$;
these are presumably related to critical slowing down near the
magnetic phase transition.
We found that an accurate value for $T_c$ was most conveniently
obtained by computing several values of $m$ in the range
$0.15 \lesssim m \lesssim 0.3$
$(0.02 \lesssim m^2 \lesssim 0.1)$ and finding $T_c$ by
fitting to $m^2(T) = \alpha (T_c-T)$ with $\alpha$ and $T_c$
fit parameters. 

In previous work \cite{Millis95a,Millis95b} we had also
used an alternative method (which we termed the {\it
projection method}) based on the observation that
by choice of an appropriate local spin reference frame
one may map the model into one of spinless fermions moving
in a lattice with a spatially varying hopping determined
by the local spin orientations.  We further argued that within
mean field theory one could approximate this hopping by
$t(m)=\sqrt{(1+m^2)/2}$, thereby simplifying the
problem to one of spinless fermions, with hopping
$t(m)$, coupled to phonons.  Finally we argued that one could 
construct a mean-field magnetic free energy by combining the
m-dependence of the free energy of the auxiliary problem
with the entropic term from the conventional mean field theory
for Heisenberg spins.  This procedure leads to
a $T_c(lambda)/T_c(0)$ almost identical
to that shown in Fig. 2; however the {\it projection method}
 $T_c$ is lower than those shown in Fig. 2.  For
example, the direct-integration $T_c$ at $\lambda=0 \; \mu=0$ is
$0.1t$, much less than the $0.17t$ shown in Fig. 2.  A numerical
error originally led us to believe the $T_c$'s of the
two approaches coincided.  The discrepancy may most easily be
understood by expanding $F=-Tln[Z_{LOC}]$ to order $a_{1n}^2$.
The result is a quadratic form 
$\delta F \sim \sum_{mn}a_{1n}\Lambda_{nm}a_{1m}$.
For example, at $g=0$, $\Lambda_{mn}=\delta_{mn}(1-1/3(a_{0n})^2)
+\frac{2}{3a_{0n}a_{0m}}$.
$T_c$ is the temperature at which ${\bf \Lambda}$ first has a 
zero eigenvalue.  The prjection method  result
corresponds to setting $\Lambda_{mn}=\delta_{mn}(1-1/3(a_{0n})^2)$
and $a_{1n}=constant$; in other words it produces a lower
$T_c$ because it does not permit an optimal
choice of $a_{1n}$.  We have therefore not used the direct
integration method in this paper.  We note, however, that the
projection method provides a transparent and physically appealing
motivation for the result, found also in the detailed calculations
presented below,
that the $T_c$ is controlled
by the kinetic energy at $T_c$, is borne out
by our detailed solution of the model.

\section{Qualitative Behavior}
In this section we discuss the qualitative behavior
of the solutions of Eqs. \ref{eq:MF2}.
Much of the behavior is similar to that found in I.
The new feature is the physics of double
exchange, which is expressed via $b_1$, 
via the angular integral  and via the
factors of $\frac{1}{2}$ on the right hand side
of Eqs. \ref{eq:MF2}.

At $T \rightarrow 0$, the $\theta$ integral is dominated
by the regime $\cos \theta =1$, so
$b_1 = b_0 -(\omega + \mu)$. From Eq. \ref{eq:GLOCUP} 
one sees that at $T=0$ the antialigned component
of G vanishes, while the aligned component is determined by
$(b_0 + b_1)$ which is given by an equation identical to that
considered in I.
Therefore,  all of the results obtained 
in I for the $T \rightarrow 0$
limit hold also here.
At $T>T_c$, there is no long range magnetic
order.  Thus $b_1=0$, there is no $\theta$-dependence
and $b_0$ is given by an equation which differs by a factor
of $\frac{1}{2}$ from that treated in I.

Further insight into the quantity $b_1$ may be gained from the
$\lambda =0$ limit.
At $T=0$ and $\lambda =0$ the quantity $b_0 + b_1$ is found
from Eqs. \ref{eq:MF} to be
\begin{equation}
b_0 + b_1 = \frac{1}{2} \left[ \omega + \mu -i
\sqrt{4-(\omega + \mu)^2} \right]
\label{eq:NONINT1}
\end{equation}
This is precisely the usual non-interacting
solution:
$Im G_{LOC} \neq 0$ in a semicircular band of full
width $4t$.
In the present conventions, the fermi level is at
$\omega =0$ and for $\lambda =0$ the maximum of the
spectral function is at $\omega = - \mu$.
The self-energy for this solution vanishes.

At $T>T_c$ and $\lambda =0$, $b_1 =0$ and
\begin{equation}
b_0 = \frac{1}{2} \left[ \omega + \mu -i
\sqrt{2-(\omega + \mu )^2} \right]
\label{eq:NONINT2}
\end{equation}
Here $Im G_{LOC} \neq 0$ in a semicircular band of full
width $2\sqrt{2}t $: the fact that neighboring
spins are uncorrelated has reduced the bandwidth, and thus
the kinetic energy, by a factor of
$\sqrt{2}$.
This may also be seen by a direct evaluation
of $K$ from Eq. \ref{eq:K}.
Further, the self energy is
\begin{equation}
\Sigma (\omega) = -b_0 = \frac{i}{2}
\sqrt{2-(\omega + \mu)^2} - \frac{1}{2}
(\omega + \mu)
\label{eq:sigmanonint}
\end{equation}
and has a non-zero imaginary part at the fermi
surface $(\omega =0)$, corresponding physically
to scattering by spin disorder.
However, this scattering is not too strong.
From Eqs. \ref{eq:sigmanonint},
\ref{eq:D} one finds that the product of the imaginary
part of the self energy and the density of states at the
Fermi level is
$(2- \mu^2)/ \pi$.  This number is
rather less  1, and implies a mean free path 
longer than $p_F^{-1}$.
This spin disorder scattering decreases as $T$ is decreased
below $T_c$.

The model with $\lambda =0$ was studied in the
dynamical mean field method by Furukawa \cite{Furukawa94},
who obtained Eq. \ref{eq:sigmanonint}. Furukawa
also used a method
he referred to as solving the equations at constant magnetization
to produce an interpolation formula describing the
temperature dependence of $\Sigma^{\prime\prime}$ for
$0 \leq T \leq T_c$.
We believe these results are similar but not quite
equivalent to those we obtain by solving the equations
\ref{eq:MF2} directly.
However, the minor differences between Furukawa's results and ours
are not important.
The main point is that the scattering at $T >T_c$ predicted
by this calculation is much too small to explain the data.

One may calculate $T_c$ at zero coupling by linearizing
the second of Eqs. \ref{eq:MF2} in $b_1$.
One finds that $T_c(\mu)$ is given by the solution of
\begin{equation}
T_c(\mu) = -\int_{-\sqrt{2}}^{\sqrt{2}}\frac{d\omega}{\pi}
f((\omega-\mu)/T_c(\mu))
\frac{\omega\sqrt{2-\omega^2}}{8/3-\omega^2}
\label{eq:Tcnonint}
\end{equation}
where f is the fermi function.

We now return to the issue of the effects of the electron-phonon
coupling.
At $T=0$ the mean field equation is identical to that considered
in I. From this work we learn that there are three regimes:
{\it weak coupling}, in which 
$lim_{T \rightarrow 0} \bar{x}^2 (T)=0$,
$lim_{T \rightarrow 0} \rho(T)=0$ and
$d\rho/dT |_{T=0} \sim \lambda$; {\it intermediate coupling},
where $0 < lim_{T \rightarrow 0} \bar{x}^2(T) < \bar{x}_c^2 \sim 1$,
$0 < lim_{T \rightarrow 0} \rho(T) < \infty$ and
$d\rho/dT |_{T=0}$ may have either sign, and {\it strong
coupling}, where $\bar{x}_c^2 < lim_{T \rightarrow 0} \bar{x}^2(T)$
and $lim_{T \rightarrow 0} \rho (T)= \infty$.
Here $x_c$ is the value of frozen in lattice
distortion above which a gap appears in the electron
spectral function.
In the strong coupling regime one may think of the electrons
as being localized as polarons.

Another crucial result of I is that the transitions
between the different regimes are controlled by the values of an
effective coupling determined by the
ratio of an electron-phonon energy to a kinetic energy.
As we have seen, the kinetic energy is temperature dependent
because of double exchange; thus as temperature is
varied the behavior of the model may change from ``metallic''
$(d\rho/dT >0)$ to insulating ($d\rho/dT < 0)$.
As T is decreased below $T_c$ there are two effects 
causing a decrease in the resistivity: the spin scattering
freezes out and the effective electron-phonon coupling weakens.

\section{Half-filling}
In the section we present and discuss results of numerical
calculations for the particle-hole symmetric ($n=1$) case.
We begin with ferromagnetic $T_c$ shown in Fig. 2.
One sees that $T_c$ decreases with increasing $\lambda$;
the variation is particularly rapid in the region
$\lambda \approx 1$ which is shown below to be the critical value
at which the model goes from metal to insulator. 

For $n=1$ and all $ \lambda$ we verified that the transitions
were second order by comparing the $T_c$ obtained in this
manner to the $T_c$ obtained by determing the temperature
at which the non-magnetic state becomes linearly unstable.
We also checked for metastability at various $n$ and $\lambda$
by starting our iterations with saturated magnetization
$(b_1 = b_0 - (\omega + \mu))$ and with very small $b_1$, and
verifying that both initial conditions converged to the
same solution.
The magnetic transition was always found to be second order.

It is interesting to
compare our results to those of Ref. \cite{Roder96}.
The method used by these authors to treat the magnetic fluctuations
is very similar to the "projection method" discussed
at the end of section II.  We found that the method
did not give an accurate value for $T_c$ but did reproduce the
coupling dependence well.  Ref \cite{Roder96} used a model
with one orbital per site; we should therefore compare their results
for $n=1/2$ to ours for $n=1$.  Their quantity 
$\epsilon_p=\lambda_{JT}^2
/2K$ corresponds precisely to our $\lambda$;
the factor of two comes from the orbital degeneracy as
explained in our previous paper I.  As far as
can be determined from Fig 1 of ref \cite{Roder96}, their
calculated $T_c(\epsilon_p)/T_c(0)$ agrees very
well with our $T_c(\lambda)/T_c(0)$.  The correspondence is
interesting because the calculation of ref \cite{Roder96} was
done with quantum phonons with the rather high frequency
$\omega=0.5$ (units not specified, but presumably
set by the electron hopping, $t$).  
This supports our 
claim that quantum
effects are not important at temperatures of the order of $T_c$.
We believe the results presented in \cite{Roder96} for the doping
dependence of $T_c$ are not physically relevant because they
are based on a model with one orbital per site, which therefore
has no kinetic energy at $n=1$.  They argue that the one-orbital model
is justified by the existence of the Jahn-Teller splitting.
Our results show that this is not the case.  We discuss the 
physics of the doping
dependence of $T_c$  further in the conclusion.

We now turn to
the temperature dependence of the resistivity,
shown in Fig. 3.
The curves display kinks at the ferromagnetic $T_c$.
The resistivity drops as T is decreased
below $T_c$ both because the magnetic contribution to the
scattering begins to decrease at $T_c$ and
because the effective electron-phonon
interaction becomes weaker. From these curves we may distinguish
``metallic'' ($d\rho/dT > 0)$ and ``insulating''
$(d\rho/dT < 0)$ regimes.
At $T > T_c$ $\lambda =1$ marks the
boundary between metallic and insulating regimes;
at $\lambda =1$, $d\rho/dT=0$.
For $T<T_c$ the crossover occurs at the somewhat larger
$\lambda \sim 1.15$.
The difference in the critical $\lambda$ required to
produce insulating behavior reflects the effect of
spin alignment on the electron kinetic energy.
We also note that although it is difficult to
perceive on the logarithmic scale used in Fig. 3,
for $1.08 < \lambda < 1.15$,
$lim_{T \rightarrow 0 \rho (T)} = \rho_0$
is neither zero nor infinite.
For $\lambda$ sufficiently close to 1.15,
$lim_{T \rightarrow 0} \frac{d \rho}{dT} <0$.
Similar behavior was discussed at length in I.

Figure 4 shows the temperature dependence of the mean square
lattice distortion $\bar{x}^2$. From this one may 
distinguish the low-T {\it weak, intermediate} and 
{\it strong coupling} 
regimes, based on the $T \rightarrow 0$ limit of $r(T)$.
The regimes were discussed at length in I.
Roughly, in {\it weak coupling} $r(T=0)=0$; in
{\it intermediate coupling} $0 < r(T=0) <1$ and
in {\it strong coupling} $r(T=0) >1$.
In intermediate coupling there is a frozen-in lattice distortion
which affects the $T=0$ physics but is not large enough
to open a gap; for strong coupling
the distortion is large enough
to open a gap and cause insulating behavior.
For $r^2>0.25$, $lim_{T \rightarrow 0}
d\rho / dT < 0$ even though if $r^2 <1$,
$lim_{T \rightarrow 0} \rho (T)$ is finite.

The effects of double exchange may be seen 
in Fig. 4.  $T_c$ is visible as a kink on each curve.
At $T>T_c$, $dr^2 /dT$ decreases, implying a stronger 
electron-phonon coupling.
For $\lambda >0.9$ the $T=0$ values obtained by
extrapolating the $T>T_c$ curves to 0 are non-zero,
and are higher than the actual $T=0$ values,
because the reduction of kinetic energy due to
spin disorder has effectively made the electron-phonon coupling
stronger.
Note that the $T=0$ extrapolation of $T>T_c$ portion of the
curve corresponding to $\lambda =1.05$ is about $r^2 = 0.95$.
This is slightly less than the critical value $r_c^2=1$ found in
I to mark the boundary between finite and infinite 
$\rho(T=0)$ at $n=1$. From this 
we would infer at $T>T_c$, $d\rho /dT$ changes
sign at $\lambda \gtrsim 1.05$,
as indeed is seen in Fig. 3.

The curves presented in Fig. 4 show $r^2$ in arbitrary units.
To estimate the magnitude of the effect in $Re_{1-x}A_xMnO_3$ 
we note that in $LaMnO_3$ each O-ion is displaced $\approx 0.15 \AA$
from its ideal perovskite position \cite{Ellemans71}.
The estimates obtained in \cite{Millis96a} imply $\lambda \approx 1.3-1.5$
in that material; thus $r^2 = 3$ in fig. 4 corresponds to
an rms displacement of an O-ion of about $0.15 \AA$.

We next consider the temperature dependence of the kinetic energy
shown in Fig. 5.
At $\lambda =0$ the kinetic energy changes by about
$1/ \sqrt{2} =30$\% between
$T=0$ and $T=T_c$, and has a weak T dependence at
$T>T_c$.
For $\lambda =0.71$ the kinetic energy changes
between 0 and $T_c$ by a somewhat larger amount;
for $\lambda = 1.11$, by still larger amount,
for $\lambda =1.29$, yet larger.
These changes come from the previously discussed interplay
between double exchange and electron-phonon coupling.
As T is increased from zero, the spins disorder.
This reduces the electron kinetic energy and permits the
electron-phonon coupling to further localize the electrons,
reducing their coupling yet more, etc.
We also note that we found
the ratio between $T_c$ and the kinetic
energy at $T_c$ to be the same within a few percent for
all $n$ and $\lambda$ studied.  For $n=1$ this can be seen by
comparing Figs 2 and 5.

These arguments also explain the magnetic field
dependence of the resistivity.
Increasing the field aligns the spins, increases the
kinetic energy, and decreases the effective electron-phonon
coupling, leading to a large change in resistance
as shown in Figs 6a and 6b.
When the decrease in effective electron-phonon coupling
tunes the model across the ``metal''-``insulator''
transition, as in Fig. 6b, 
the  magnetoresistance is particularly large.

Further insight into the interplay of double
exchange and localization comes from the
optical conductivity shown in Fig. 7.
Panel 7a shows $\sigma ( \omega )$ at different T
for the weak coupling $\lambda = 0.7$.
At low T $\sigma$ has approximately the Drude form
$\sigma(\omega) = \Gamma/(\omega^2+\Gamma^2)$
with scattering rate $\Gamma \sim T$
as expected from classical phonons.
As T is increased through $T_c$ the Drude peak broadens
and acquires a T-independent part, due to spin scattering.
Panel 7b shows $\sigma$ for the moderate coupling $\lambda =1$.
At low T, $\sigma$ has the Drude form; as T is increased a broad
peak develops;
this is due to transitions between the two Jahn-Teller
split levels.
It is broad because the phonon coordinate is strongly
fluctuating, so the level position is not well defined.
As T increases beyond $T_c$ the peak broadens
almost to indistinguishability.
Note also that as T is increased, the optical
spectral weight $\int d \omega \sigma ( \omega )$ decreases,
reflecting the increasing localization of electrons by phonons.
In models such as the present one which do not have Galilean
invariance and involve only a limited number of orbitals,
the f-sum rule spectral weight is not constant
and is indeed proportional to the kinetic energy \cite{Millis90}.
Panel 7c shows that at a stronger coupling
$\sigma$ does not have the
Drude form, and the peak is already evident at $T_c/2$.
Note that the maximum in $\sigma$ has moved
to a slightly higher frequency.
Recently
published data of Okimoto et al. on
$La_{1.825}Sr_{.175}MnO_3$ \cite{Okimoto95} are
similar to the curves shown in Fig. 7c, although our
use of classical phonons means that we cannot obtain
the very narrow Drude peak found at low T.
Panel 7d shows $\sigma$ at the still stronger
coupling $\lambda =1.15$ where the model
has a large frozen-in lattice distortion even at $T=0$.
The $\sigma$ has an insulating appearance,
above and below $T_c$, but as T is decreased the peak
in $\sigma$ shifts to a lower freqency
and grows in intensity, reflecting the effectively weaker coupling.
The nonmonotonic behavior of 
the dc conductivity is not reflected in
$\sigma (\omega)$ at $\omega \gtrsim 0.5$.
The curves in Fig 7d resemble data recently
obtained by Kaplan et. al.  on $Nd_{0.7} Ca_{0.3}MnO_3$
\cite{Kaplan96}.

We will discuss the physical interpretation of
$\sigma (\omega ,T)$ in more detail in the next section
and in the conclusion.
Here we note that in the strong-coupling regime the
two d-states on a site are split.
The peak in the optical conductivity corresponds roughly to a
transition in which an electron moves from an occupied orbital
on one site to an unoccupied orbital on an adjacent site.
In our classical approximation, ``Franck-Condon'' transitions
involving also emission or absorption of a phonon cannot occur
at all.  In a more realistic model such effects
would e.g. increase the low-frequency tails by a small
amount.
The width of the peak in $\sigma (\omega )$ is determined by the
broadening of the localized states due to electron hopping and by
thermal broadening, which leads to a range of lattice distortions and
thus to a range of splittings.
\section{Different Dopings}
In this section we present and discuss results of numerical calculations
for the particle-hole asymmetric case $n \neq 1$.
As discussed in I (see especially Figs. 10, 11), at $n \neq 1$ 
in the strong coupling limit the spectral function has a
three-peaked structure.  The outer two peaks represent the Jahn-Teller-split $e_g$ levels on occupied sites, and occur also for
$n=1$.  The middle peak comes from  unoccupied sites, 
on which there is no Jahn-Teller splitting.
These states tend to fill in the gap created by 
the Jahn-Teller splitting and
mean that stronger coupling is required to obtain insulating behavior
at $n \neq 1$ than at $n=1$.
Further, in the strong coupling limit the temperature dependence
of physical quantities is determined by the energy difference between
filled and mid-gap states; thus at fixed Jahn-Teller splitting the
activation gap for physical properties is much less at $n \neq 1$
than at $n=1$.
Note also that $T_c$ is controlled by the 
electron kinetic energy which
is in turn controlled by the Jahn-Teller splitting.
Therefore, in the strong coupling limit at fixed $T_c$ 
the activation gap characterizing
the $T >T_c$ resistivity is much larger at $n=1$ than at $n \neq 1$.

This physics is immediately apparent in the resistivity curves
for $n=0.75$ and $n=0.5$ shown in Fig. 8.
Comparison to Fig. 3 shows that much stronger couplings are required 
to obtain a $d \rho /dT <0$ for $n=0.75$ than for $n=1$ and
stronger couplings yet are required 
for $n=0.5$.   The smaller
value of the activation
gap relative to $T_c$ means that the resistivity rises less
before the behavior changes at $T_c$
for $n \neq 1$ than for $n = 1$.
For $n=1$, $\lambda =1.1$  we found an order-of-magnitude
rise in $\rho$ as $T$ is decreased to $T_c$, and 
we found metallic behavior below
$T_c$.  At $n \neq 1$ it is difficult to produce much
of an up-turn in $\rho$ at $T> T_c$ for parameters such
that the model is metallic at $T=0$.
This physics is due to the particular (Jahn-Teller)
form of the electron-phonon
coupling we have chosen to study.

The same arguments mean that it is not possible to get as large a
magnetoresistance at $n \neq 1$ as at $n=1$.
Figure 9 shows the temperature dependence of the resistance
for several different coupling strengths and magnetic fields.
To get even a moderately large effect one must choose a very
strong coupling, such that the model is insulating for both $T >T_c$
and $T < T_c$.

Finally, Fig. 10 displays the temperature dependence of the optical
conductivity at $n=0.75$ and 
moderate ($\lambda = 1.29$) and strong
($\lambda =1.49$) coupling, and compares this to the momentum-integrated
spectral functions.
One sees by comparing energies that at strong coupling the two
maxima in the conductivity may be associated with transitions from
the lowest peak in the spectral function (representing occupied orbitals
on occupied sites) to the middle feature (representing unoccupied
orbitals on unoccupied sites) and to the higher feature (representing
unoccupied orbitals on occupied sites).
At $n=1$ the middle peak in $A$ is absent and $\sigma$ 
has only one peak, as seen in Fig. 7d.
Of course, the on-site d-d transition is not optically active:
the calculated conductivity involves electron motion from one site
to another, and for this reason $\sigma$ is not simply given by a
convolution of two local spectral functions.
One may see this in Fig. 10.
The central peak in $A( \omega )$ has less area than the upper one, yet
the lower peak in the corresponding $\sigma (\omega )$ is the larger.
This may be understood from the above arguments:
a transition from the lower to the middle peak of $A(\omega )$
necessarily involves moving an electron from one site to the other,
but some of the transitions from lower to higher are on-site transitions
which do not contribute to $\sigma$.

Note also that the $T = 2T_c$ spectral function
shown in Figs 10b,d has a sharp minimum
at $\omega + \mu =0$.
This is a consequence of the  fact, discussed in I, that the
probability of a small-amplitude lattice distortion is small because of
the $xdx$ measure, and decreases as $T$ increases, due to the shift
to higher $\langle x^2 \rangle$ of $P(x)$.
As can be seen in the corresponding optical conductivity
curves in Figs 10a,c, this
minimum is of little significance for other physical quantities.
\section{Conclusion}
We have used the ``dynamical mean field'' approximation to solve
a model of electrons ferromagnetically coupled to classical spins and
Jahn-Teller coupled to localized classical oscillators. In a companion 
paper (I) we considered electron-phonon coupling
in a variety of models without double exchange.
The results presented in Section IV for the half-filled case bear
a striking resemblance to data for the ``colossal magnetoresistance''
materials $Re_{1-x}A_xMnO_3$ in the $0.2 < x < 0.5$ regime where
the ground state is metallic.  We believe the
agreement supports the idea that the
important physics of  $Re_{1-x}A_xMnO_3$  involves the
interplay between a strong electron-phonon coupling and the
``double exchange'' effect of magnetic order on the electronic
kinetic energy.  Specifically, the $\rho(T)$ curves shown in
Fig. 3 are very similar to those shown, e.g in \cite{Tokura94,Hwang95}.
Varying the electron-phonon coupling produces changes very similar
to those found experimentally by varying $x$ and the
constituents $Re$ and $A$.  The magnetic field dependences
shown in Fig 6 also bear a striking resemblance to data.
Fig 6b looks very like Fig. 2 of \cite{Schiffer95}, while 
Fig 6a resembles magnetoresistance data which would be observed
for $La_{1.6}Sr_{0.4}MnO_3$. (It should be noted,
however, that the fields used to produce our curve, although very 
small compared to microscopic energies, are larger than experimental
fields by a factor of about 5).  The variation of the rms lattice
distortion shown in Fig 4 has been observed
via measurements of the $e_g$ component of the oxygen
Debye-Waller factor \cite{Dai96,Radaelli96}.  

Further, optical conductivity 
data of Okimoto et. al. \cite{Okimoto95}
on $La_{1.825}Sr_{.175}MnO_3$ bear a strong qualitative
resemblance to Fig 7b, while data obtained by Kaplan et. al.
\cite{Kaplan96} strongly resemble Fig 7c.  As noted in sections IV and
V, in our interpretation the higher frequency peak in $\sigma(\omega)$
is due to transitions between levels split by an electron-phonon
coupling.  Okimoto et. al. interpreted the higher peak
differently, attributing it to transitions from an inital
$e_g$ state aligned to the core spin to a final $e_g$ state
antialigned to the core spin.  They argued that their identification
was supported by the fact that in their data the higher peak was only
visible at $T > T_c$, and vanished at low T when all spins were aligned.
However, the data of Kaplan et. al. demonstrate that in some samples
the high frequency peak does not vanish below $T_c$ and indeed
grows in oscillator strength as T decreases.  This rules
out the interpretation of Okimoto et. al., at least for that
sample.

The detailed qualitative agreement between data and our model
leaves little doubt that we have identified the important 
physics governing the $Re_{1-x}A_xMnO_3$ materials.  However,
several very important issues remain unresolved.  One concerns
the origin of the experimentally observed material and
doping dependence of the results, which are {\it modelled}
in the $n=1$ calculations by varying the electron-phonon
coupling.  Another is the degree of "fine-tuning"
of parameters required.  A third concerns the effects of 
omitted interactions and a fourth is that,
as shown in Section V, computations at different electron
concentrations $n \neq 1$ agree much less well with data.
In the remainder of this paper we present a qualitative
discussion of all of these issues, which, we argue, are
closely related.

We begin with the $n \neq 1$ calculations.
We showed that the differences between the $n=1$ and $n=0.75,0.5$
results are due
to the presence, for $n \neq 1$, of mid-gap states in the spectral
function (shown, e.g., in Fig. 10).
These mid-gap states occur because we used a particular form of
electron-phonon coupling, namely a Jahn-Teller coupling which splits the
d-state degeneracy on a site if there is one electron on the site,
and does nothing otherwise.
In ref \cite{Roder96} it is argued that the existence of the Jahn-Teller
coupling justifies a model involving only one orbital per site.
The results presented here suggest that this is  oversimplified,
because it does not take in to account the mid-gap states.
A model with only Jahn-Teller coupling does not suffice.
However,
results presented in I strongly suggest that 
if the model were extended in
a way which moved both the upper peak
and the mid-gap states up in energy, then the model would
become effectively a single-orbital model and results for
$n \neq 1$ would much more closely resemble those obtained for $n=1$.

One omitted piece of physics which will have precisely this
effect is the breathing-mode distortion of the oxygen
octahedron around an Mn site.  The breathing mode couples
to charge fluctuations on the Mn site.  This coupling is likely
to be at least as strong as the Jahn-Teller coupling, as
may be seen from the following argument: 
the Jahn-Teller coupling is due to the dependence of the force exerted
on an O ion on the orbital occupied by the outer-shell d-electron.
Whatever its magnitude, this force is unlikely to be larger
than the force created by simply removing that d-electron, and making
an unbalanced charge.
The breathing-mode coupling was recently argued to be important
for the small-x structural phase boundary \cite{Millis96a}.
To understand the effects of the breathing mode, consider again
Fig. 10.  The central peak in the spectral function depicted 
in the low-T curve in Fig 10d
gives the
states available for adding an electron to an unoccupied site.
If such a site has a breathing distortion already
present, the energy cost of adding an electron will be
increased, thus the middle feature will also move up
in energy, increasing the gap as required.

Another important piece of physics is the on-site Coulomb
interaction.  This must be strong because if it were not,
the Hund's coupling $J_H$ would not be large \cite{Varma96}.
The Coulomb interaction leads to two related effects.
One is most easily discussed by reference to the spectral 
functions and optical
conductivities shown in Fig. 10.
Now the $\omega > 0$ part of the spectral function corresponds to
states into which an electron maybe added; the upper peak thus
gives the states available for adding an electron onto a site which
already has an electron.
The Coulomb interaction must move such states up in energy,
and must similarly move up the second peak in $\sigma ( \omega )$.
If the Coulomb energy is of the order of the Hunds coupling,
then it is very likely that this effect will move the higher peak out
beyond the physically interesting energy range $\omega \lesssim 3$ eV.

The combined effect of the breathing distortion and the Coulomb
interaction is therefore to lead to a spectral function with
at most two peaks in the energy range of interest.  The
only difference between this realistic situation and the
situation encountered in the $n=1$ calculations is that
the realistic spectral function is not symmetric under the interchange
of the two peaks.  This asymmetry was shown in I not to be
important.

A second effect of a strong Coulomb interaction is to localize the
electrons.
It is likely that the observed very strongly insulating behavior of
$ReMnO_3$ is not due solely to the Jahn-Teller order, and that
$ReMnO_3$ is to some degree a Mott insulator.
Now the kinetic energy $K$ of a Mott insulator has a pronounced
doping dependence \cite{Millis90}.
For $Re_{1-x}A_xMnO_3$ one would expect $K(x)$ to increase with
$x$ for $x <0.5$.
Because, as we have argued at length, the properties of electron-phonon
models are controlled by the ratio of a coupling energy and a
kinetic energy, this will  lead to an x-dependence of the effective
coupling strength, with larger $x$ having a weaker effective coupling.
We believe that this strong x-dependence of the effective coupling
accounts for the ubiquity of the "colossal" magnetoresistance
phenomenon.  Different materials have different bare electron
hoppings and probably different electron-phonon couplings,
but in all materials the variation of the electron kinetic energy
with $x$ is large enough to sweep the effective coupling through
the critical value at some x between $0.1$ and $0.5$.

The breathing distortion may be studied via the dynamical
mean field formalism used here; one must simply
integrate over another variable in Eq \ref{eq:ZLOC}.
The on-site Coulomb interaction may also be
included, but one must  perform functional integrals
rather than simple integrals.  Monte Carlo techniques are required, 
the computational expense is greater and the accuracy is less.
Such an investigation would however be desirable.

Two other effects not included in the calculation
should be mentioned.
Quantum fluctuations of the phonons have been omitted.
As discussed in I, these will in the absence of long range order
or commensurate density lead to metallic behavior at
sufficiently low T, even in the strong coupling limit.
The neglect of the phonon momentum and quantum
fluctuations of core spins is not an important approximation
because we are primarily interested in phenomena at temperatures
of order room temperature; however if needed they could
be incorporated into the formalism. 
The neglect of intersite phonon correlations is potentially
more serious.
It is tempting to argue that they are unimportant because we
are interested in optical phonons, which are usually weakly
dispersing.
However, in the $ReMnO_3$ structure each O is shared by two Mn;
there must thus be a strong correlation between
Jahn-Teller distortions on adjacent sites.
In $LaMnO_3$ the Jahn-Teller distortions have long range
order; estimates presented in \cite{Millis96a}
suggest that in $Re_{1-x}A_xMnO_3$ the correlation length
of the Jahn-Teller distortions
is $\sim x^{-1}$ as long as the resistivity
is well above the Mott limit.
Extending the present calculations to include the 
effects of intersite
correlations is an important open problem.
It is worth addressing because in the present calculations
the correlation length is zero 
and the strong-coupling physics is of
polarons.
In the infinite correlation length limit,
the physics has to do with interband
transitions in a bandstructure defined by Jahn-Teller order.
The situation in the actual materials is presumably intermediate
between these two limits.

{\it Acknowledgements}
We thank A. Sengupta and C. M. Varma for helpful discussions and
G. A. Thomas and H. D. Drew for sharing data in advance
of publication and for helpful
discussions.
We are especially grateful to P. B. Littlewood for stimulating
our interest in the problem, collaborating in the early stages
of our work, and providing continuing advice and encouragement.
RM was supported in part by the Studienstiftung des Deutschen Volkes.
A. J. M. acknowledges the hospitality of the Institute Giamarchi-
Garnier during the early stages of this work, and of the Institute
for Theoretical Physics during the final stages of
preparation of the manuscript.
\clearpage
\begin{appendix}
\section{Analysis of Observed Resistivity}
In this Appendix we discuss the observed $T > T_c$ resistivities
of $Re_{1-x}A_xMnO_3$.
We note that the observed strong $x$-dependence suggests
that the number of active carriers is $x$.
For $x$ classical particles hopping with {\it probability} $W$
on a cubic lattice of lattice constant $a$,
\begin{equation}
\sigma = \frac{e^2xW}{3a k_BT}
\label{eq:classical}
\end{equation}
Using $a \approx 4 \AA$ as appropriate to $Re_{1-x}A_xMnO_3$ we have
\begin{equation}
\frac{\hbar W}{k_BT} = \frac{10^{-5}}{x\rho (\Omega -cm)}
\label{eq:hbarW}
\end{equation}

>From this equation one may easily see that observed resistivities,
which are typically greater than 0.01 $\Omega -cm$ at $T>T_c$ and
$x \lesssim 0.3$, and increase rapidly with
decreasing $x$ imply values of $\hbar W/k_BT$ much less then unity.
If $\hbar W/k_BT \ll 1	$, a particle has time to thermalize
before it moves, and a classical model is appropriate.
\section{Resistivity of Double Exchange Only Model}

In this Appendix we consider in more detail the resistivity of the
double exchange only model.
In this model, resistivity comes from spin disorder.
It is maximal at $T \gg T_c$ and vanishes at $T=0$.
The resistivity has been calculated, using methods which are essentially
perturbative in the amplitude of the spin disorder, by Kubo and
Ohata \cite{Kubo83} and more recently by two of us and Littlewood
\cite{Millis95a}.
The spin scattering was found not to be too strong.
As discussed in the text, similar results have been obtained using
the dynamical mean field method.
These calculations omit several physical effects
and have been questioned recently by Varma \cite{Varma96}.
In this Appendix we show that the omitted effects
are not important.

We begin by describing the omitted effects.
In the double exchange only model, the scattering is due to spin
disorder which, if the core spins are assumed to be classical, may be
treated as static scattering with the important proviso that the disorder
is annealed, not quenched.
When applied to a model with static scattering, the dynamical mean field
approximation with semicircular density of states is equivalent to the
coherent potential approximation (CPA) for the Bethe lattice
\cite{Georges96}.
The CPA neglects localization (as do the perturbative calculations
\cite{Kubo83,Millis95a}).
The lack of closed loops on the Bethe lattice
also means that  Berry phase
effects arising from particle motion 
in a spin background are
omitted.

We consider the Berry phase effects first.
In the double exchange model the hopping matrix element between two sites
$i$ and $j$ is with core spins characterized by polar
angles $(\theta_i , \phi_i)$, $(\theta_i , \phi_j)$, is
\begin{equation}
t_{ij} = t (cos \frac{\theta_i}{2} cos \frac{\theta_j}{2} +
sin \frac{\theta_i}{2} sin \frac{\theta_j}{2}
e^{i(\phi_i-\phi_j)} )
\label{eq:t}
\end{equation}
If closed loops are not important one may choose the $\phi_i$
independently on each site and recover the familiar
double exchange result
$t_{ij}= t cos (\theta_i - \theta_j)/2 $.
In general the $\phi$ factors around a closed loop produce something
like a magnetic field, which may scatter electrons.
In the limit of strong ferromagnetic correlations the phase
dependent term may be seen to be very small because all nearby sites
have very similar angles, which may be taken to be near 0.
In the limit of uncorrelated spins we may estimate the size of the
effective field by comparing the phase sensitive part of the hopping
to the phase insensitive $cos (\phi_i-\phi_j)/2$ part.
By integrating $t_{ij}$, around square placquette one finds that the
phase sensitive part is $\frac{t^4}{16} e^{2i\phi_i}$ while the phase
insensitive part is $t^4/4$.
Thus the rms deviation of the amplitude for an electron to move around
a placquette is $1/4 \sqrt{2}$ of the phase insensitive part.
This combined with the relative insensitivity of three 
dimensional physics
to closed loops suggest phase effects, while interesting, 
are too weak to
cause the observed strongly insulating behavior.

We now turn to localization.
The problem at hand concerns electrons with random hopping, which has
not received much attention.
Economou and Antoniou \cite{Economou77} have studied a Bethe-lattice
model in which the hopping amplitude $t$ has the symmetrical distribution
\begin{equation}
P_E (t) = \frac{2}{\pi t_1}.
\sqrt{t_1^2 -(t-t_0)^2}
\label{eq:P(t)}
\end{equation}
For this model $\bar{t}$, the mean value of t, equals $t_0$
and the variance
$\langle (t- \bar{t})^2 \rangle =t_1^2 /4$.
The double exchange model 
at $T>>T_c$ (so the spins are completely disordered)
corresponds to the distribution
\begin{equation}
P_{d-ex} (t) = \frac{2t}{t_D^2} \theta (t_D-t) 
\label{eq:P_d-ex}
\end{equation}
The localization effects of the double exchange 
distribution have not been determined.  We expect that 
because the most probable value is also the largest hopping,
the double exchange distribution will produce fewer
localized states than a semicircular distribution
with the same mean and variance.  Now from Eq \ref{eq:P_d-ex}
one sees that the double exchange distribution
has  mean $\bar{t}=2 t_D/3$ and  variance of
$t_D^2/18$.
Thus it should produce fewer localized states than the
model of Economou and Antoniou with $t_0=2t_D/3$ and
$t_1= \sqrt{2} t_D/3$, i.e. with $t_1/t_0=1/ \sqrt{2}$.
Inspection of Ref \cite{Economou77} reveals that at this
ratio of $t_1/t_0$, a negligible fraction of the states
are localized.  We therefore conclude that localization 
effects are not important.  
Ref \cite{Varma96} on the contrary
asserts that the double exchange
model with completely disordered spins is better modelled
by the Economou-Antoniou distribution with $t_1/t_0$ somewhat
larger than unity, so a non-negligible fraction of the
states are localized.  Drawing precise conclusions is
somewhat difficult because one result of Ref \cite{Economou77}
is that the number of localized states increases rapidly
for $t_1/t_0 >1$.  Nevertheless, we believe the estimate
$t_1/t_0 \approx 0.7$ obtained above shows that localization
effects are unlikley to be important.

\end{appendix}

\newpage
\section*{Figure Captions}
\begin{itemize}
\item[Figure 1.]
Qualitative temperature (T) - doping (x) phase digram
of $Re_{1-x}A_xMnO_3$, with magnetic phases ($F=$ferromagnet,
AF$=$antiferromagnet, P$=$paramagnet), structural phases
($JT=$Jahn-Teller order, no label$=$no order),
and transport regimes ($M=$ ``metal'', $d\rho/dT >0$,
$I = $``insulator'', $d\rho/dT <0)$ indicated.
The solid lines are magnetic phase boundaries, the heavy dashed line is the
Jahn-Teller boundary and the light dotted line is the metal
insulator crossover.
For $x > 0.5$ different physics, involving
charge ordering, is important at low T.
Different materials may have phase diagrams differing in
some details, and the magnetic and structural boundaries may
not coincide at low $T$.
\item[Figure 2.]
Dependence of ferromagnetic $T_c$ on coupling constant for
$n=1$ (heavy solid line), $n=0.75$ (light solid line), $n=0.5$
(light dashed line).
The analytic zero-coupling results are indicated by dots;
the analytic strong coupling $T_c= \frac{n}{12\lambda^2}$
results by the heavy dotted line for $n=1$.
Only for $n=1$ do the numerical calculations extend into the
strong coupling regime.
\item[Figure 3.]
Temperature dependence of resistivity at $n=1$ for
couplings $\lambda =0.32$ (lowest curve), 0.71,
1, 1.08, 1.12, 1.15, 1.20 (highest curve).
\item[Figure 4.]
Temperature-dependence of
mean-square lattice distortion for
$n=1$ and couplings $\lambda =0.71$ (lowest),
0.9, 1.05, 1.12, 1.2 (highest).
\item[Figure 5.]
Temperature (T) dependence of electron kinetic energy (K)
for $n=1$ and $\lambda =0$ (second lowest curve), 0.71, 1.12 and
1.29 (highest curve). The lowest curve corresponds to $\lambda =0$ 
in the model without double exchange.
\item[Figure 6.]
Temperature dependence of resistivity at different values of
magnetic field, $h$, for $\lambda =0.7$ (Figure 6a)
and $\lambda =1.12$ (Figure 6b).
The parameter $h$ is related to the physical field $h_{phys}$
by $h=g \mu_B S_c h_{phys}/t$.
Using $g=2$, $t=.6$ eV and $S_c=3/2$ means
$h=0.01$ corresponds to $h_{phys}=15T$.
\item[Figure 7.]
Optical conductivity, $n=1$, $T=0.02$ (light solid line),
$T=T_c/2$ (light dashed line), $T=3T_c/4$ (light dotted line),
$T=T_c$ (heavy solid line), $T_c=2T_c$ (heavy dashed line).
Panel a: $\lambda =0.71$ ($T_c =.15)$, panel b:
$\lambda =1$ ($T_c=0.10$), panel c: $\lambda =1.08$
($T_c=0.08$), panel d: $\lambda =1.15$ ($T_c=0.0675)$.
Note that in panel a the lowest T  is .025 not .02, and
$\sigma(\omega =0)$ for this curve is $21.4$.
\item[Figure 8.]
Resistivity ($\rho$) vs temperture (T) for $n=.75$
(upper panel) and $n=0.5$ (lower panel) and
couplings $\lambda =0.71$ (lowest), 1.12, 1.41, 1.49, 1.58
(highest).
\item[Figure 9.]
Magnetic field dependence of resistivity for $n=0.75$ and
$\lambda =1.12$ (lower panel), $\lambda =1.46$
(middle panel) and $\lambda =1.49$ (upper panel).
\item[Figure 10.]
Optical conductivities and spectral functions
for $n=0.75$, $\lambda =1.29$ (panels a,b) and
$\lambda =1.49$ (panels c,d).
Panel a:
$\sigma ( \omega )$, $\lambda =1.29$ and
$T=.04$ (light solid line),
$T=.061$ (light dashed line),
$T=.081=T_c$ (light dotted line) and 
$T=.162$ (heavy solid line).
Panel b:
spectral function, $\lambda =1.29$ and $T=0.04$
(solid line) and $T=0.162$ (dashed line).
Panel c:
$\sigma (\omega )$, $\lambda =1.49$ and $T=0.02$ (light
solid line), $T=.028$ (light dashed line) $T=.045$
(light dotted line), and $T=0.059=T_c$ (heavy solid line)
and $T=.115$ (heavy dashed line).
Panel d:
spectral function, $\lambda =1.49$, $T=0.02$ (solid line) and 
$T=.115$ (dashed line).
\end{itemize}
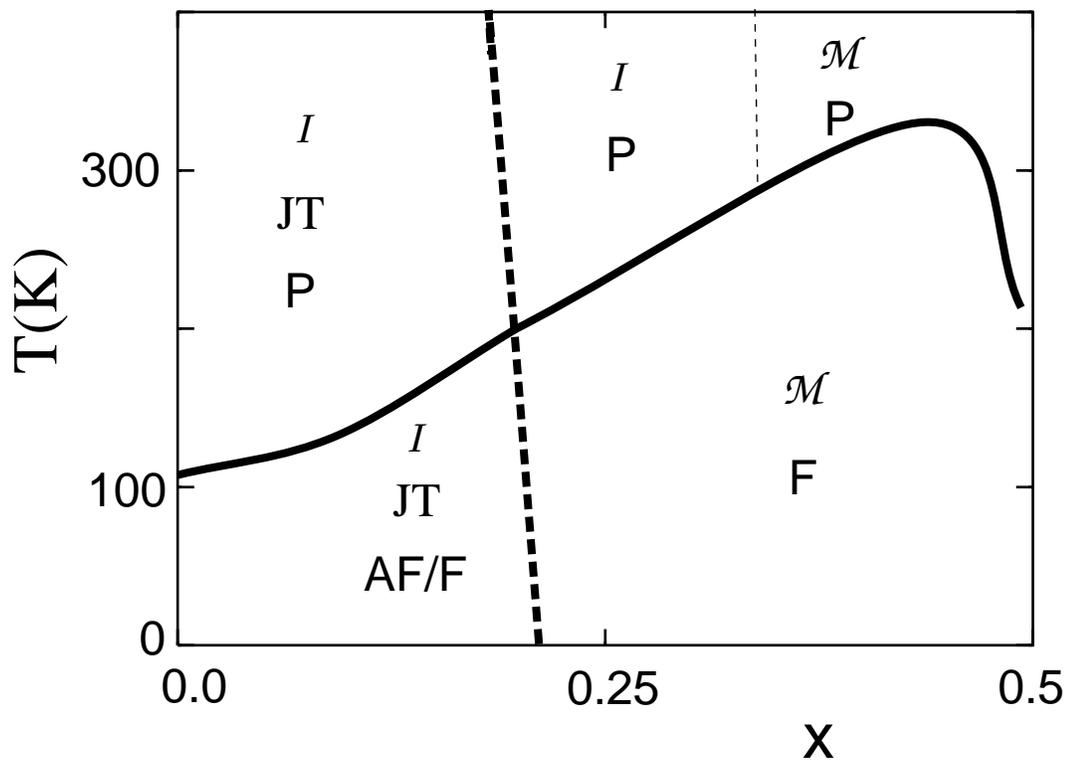
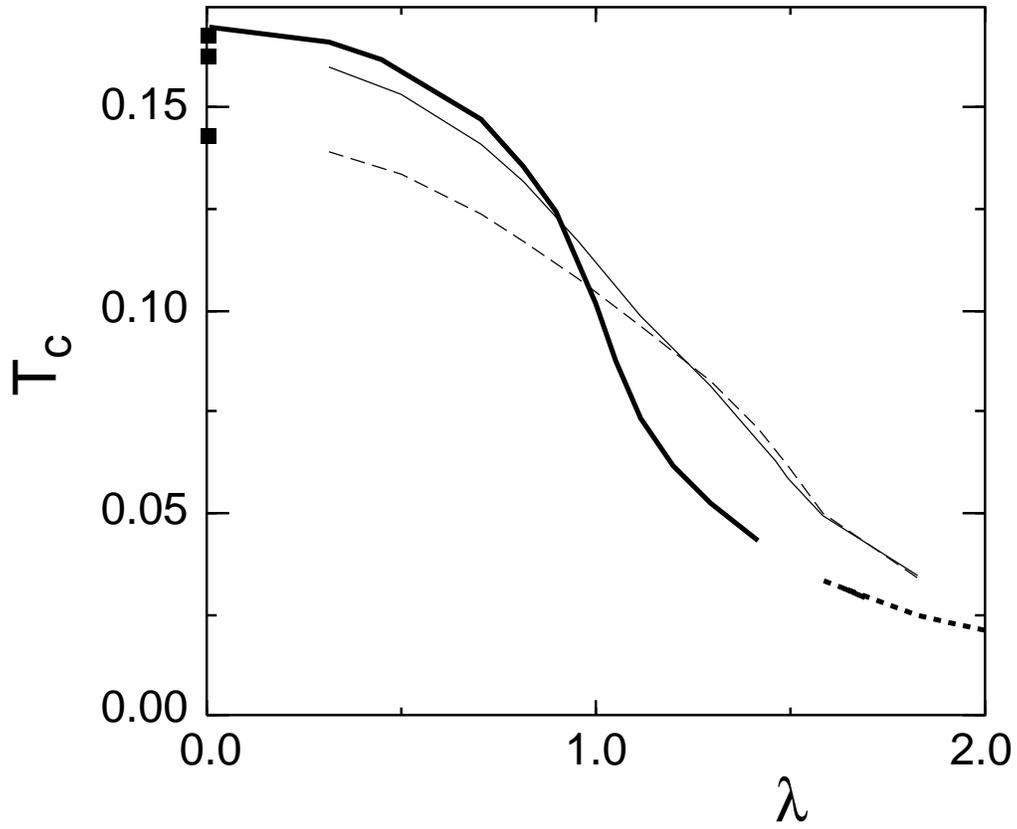
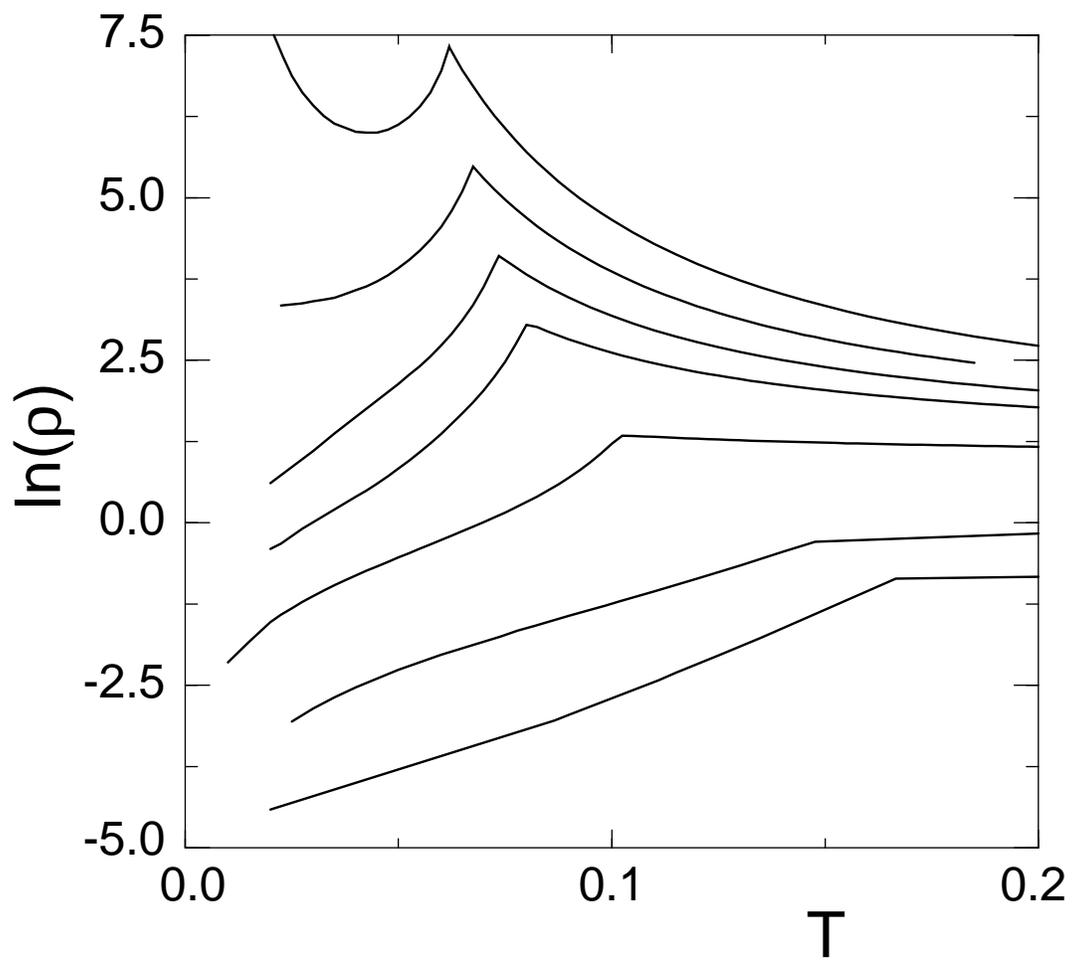
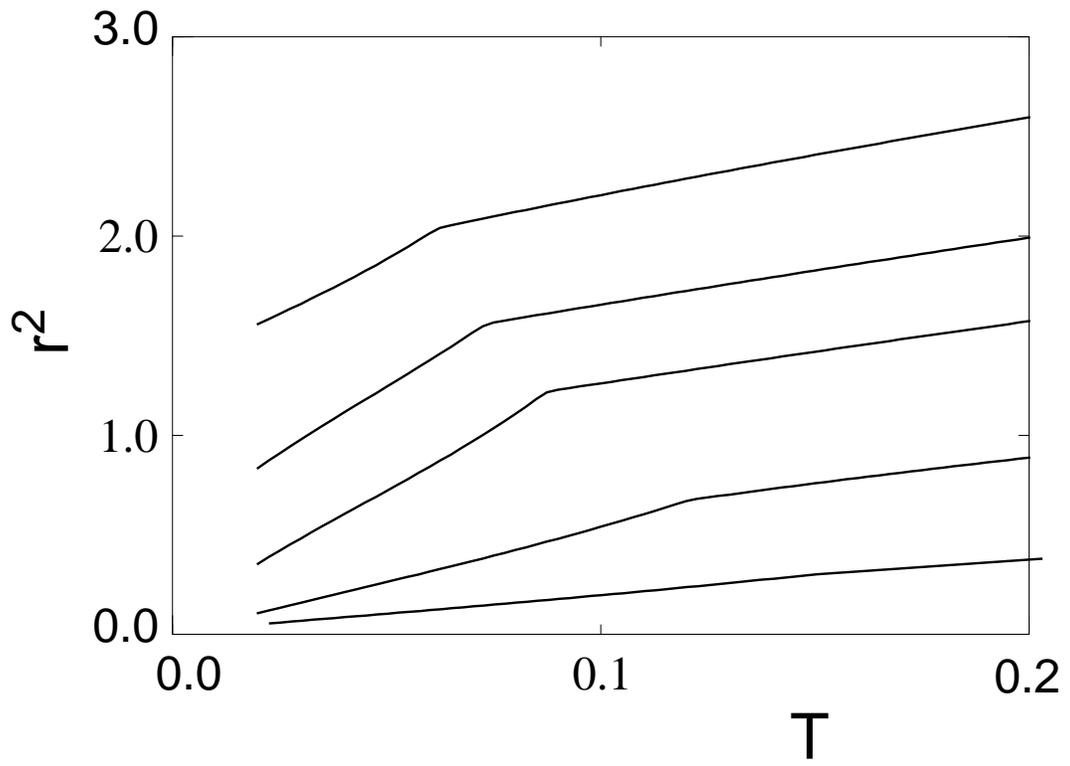
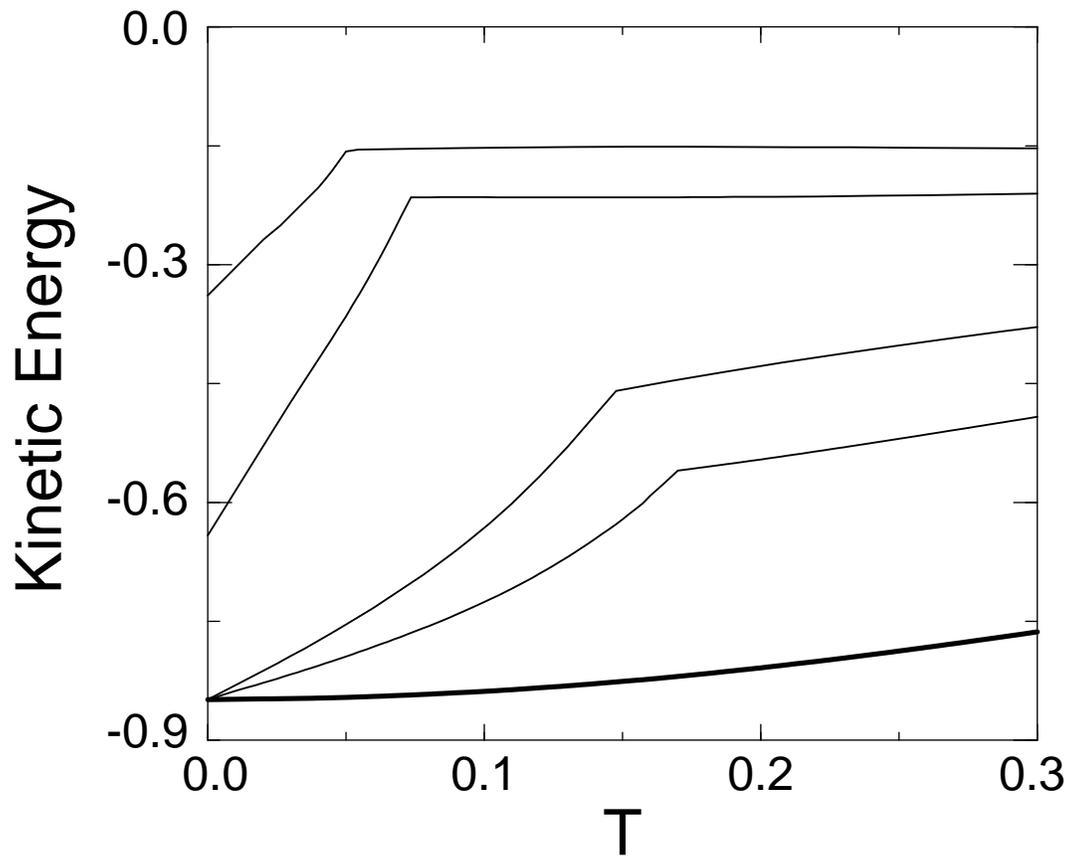
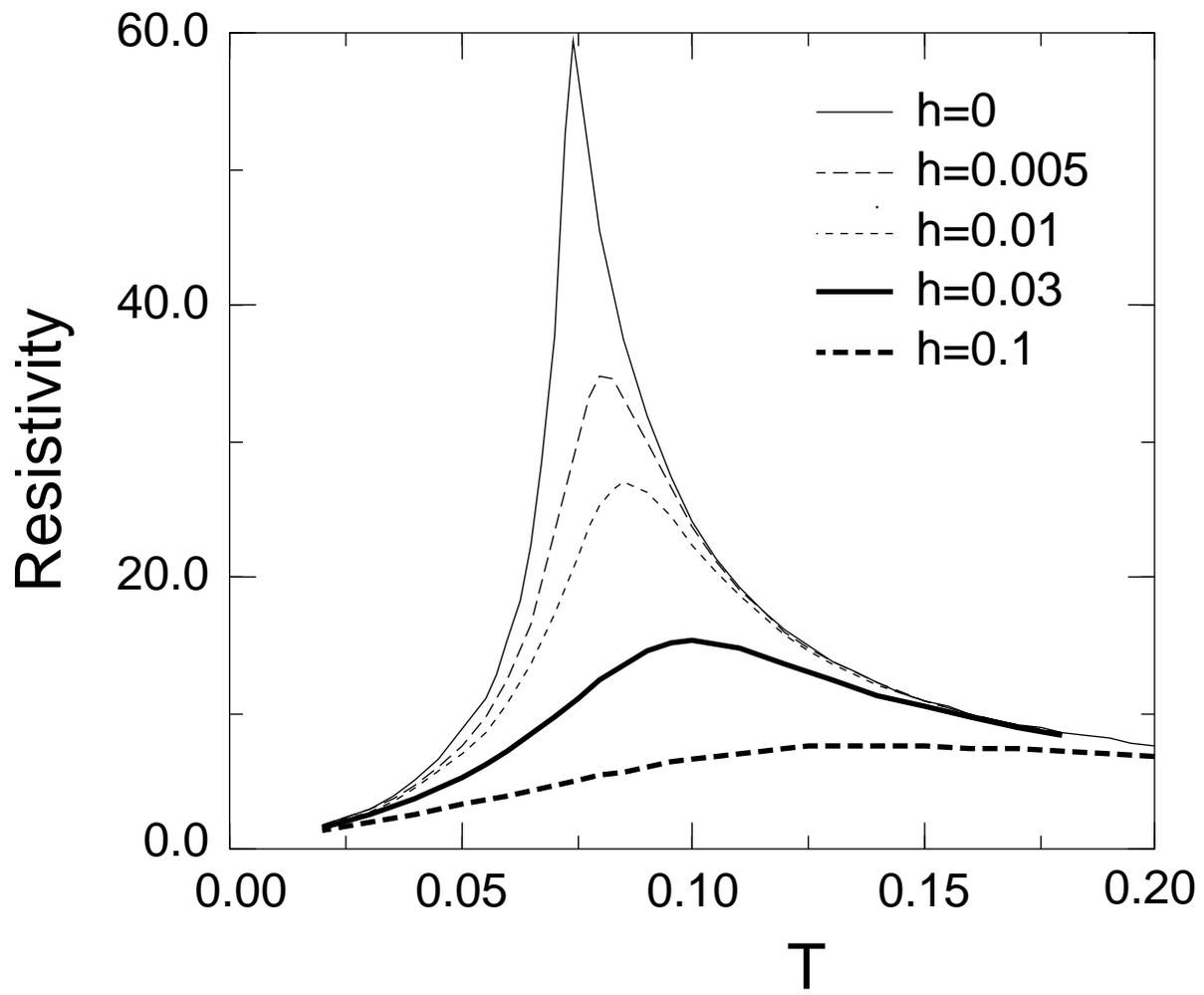
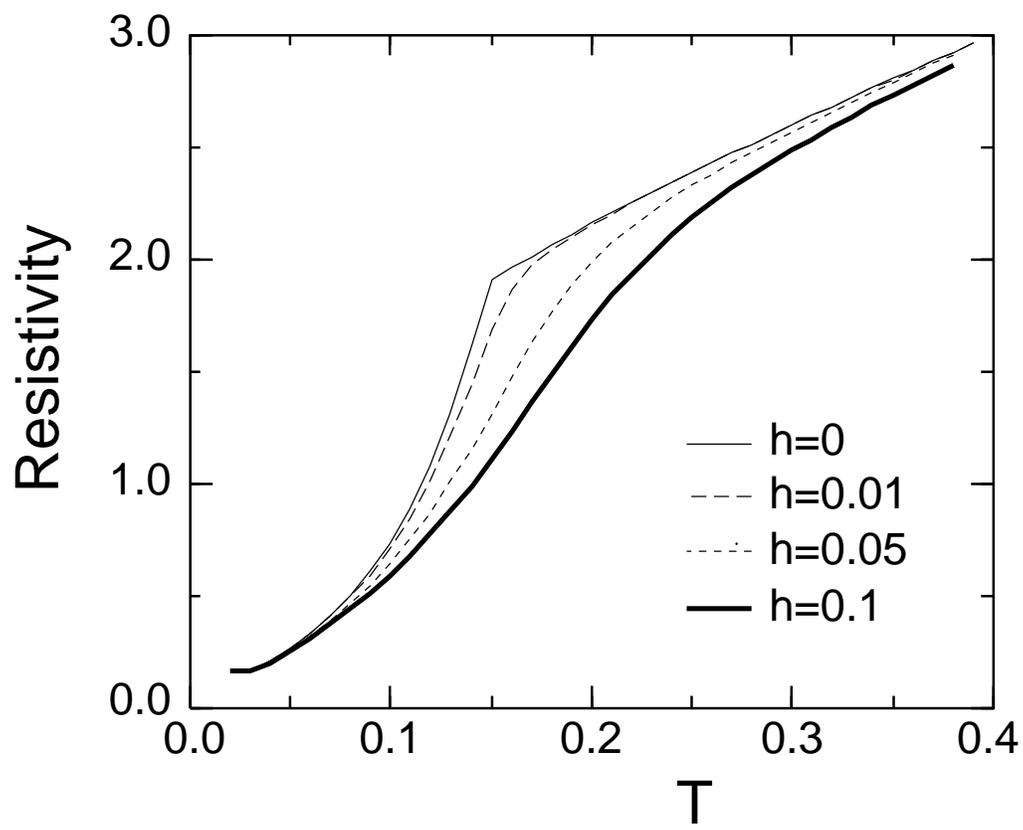
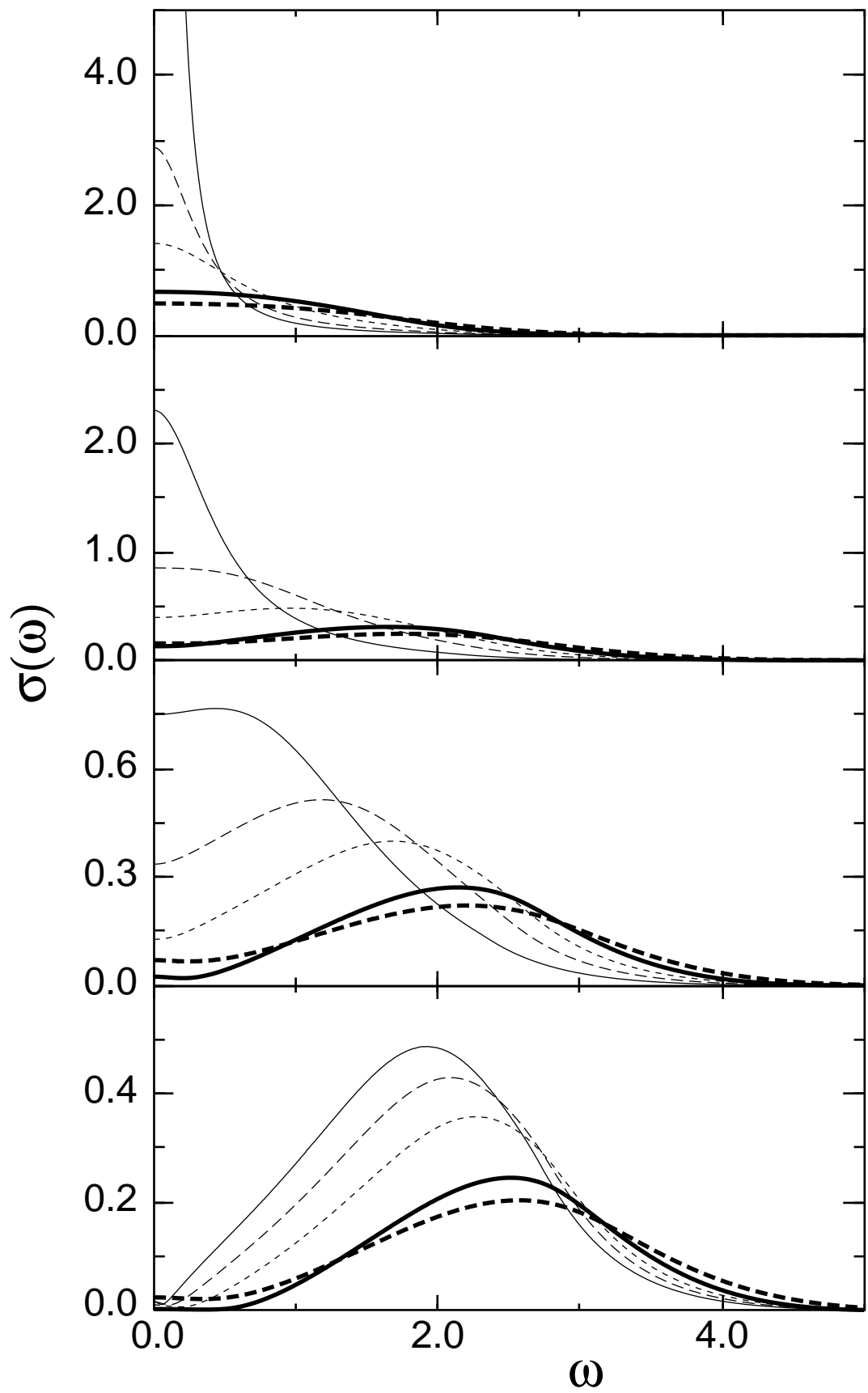
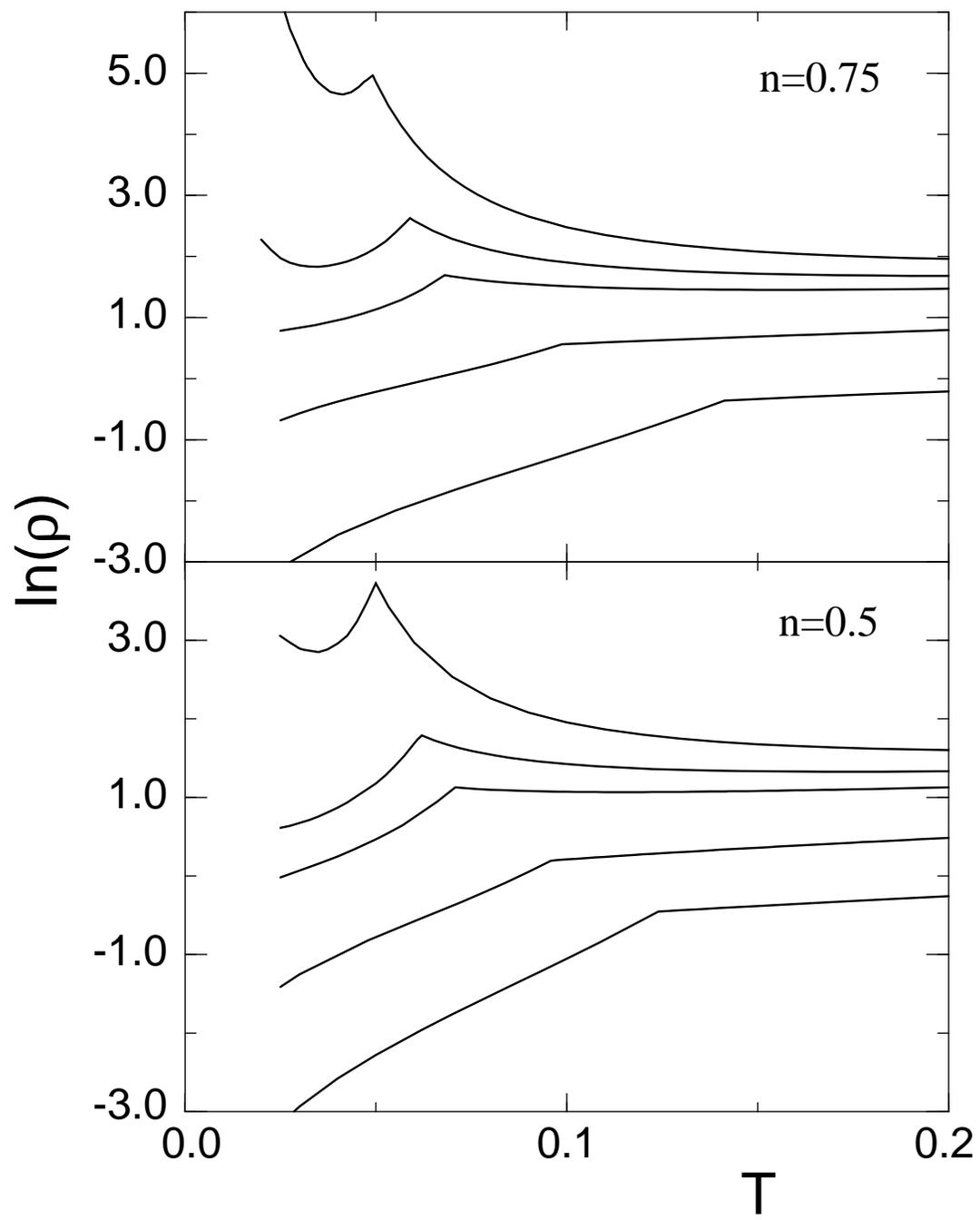
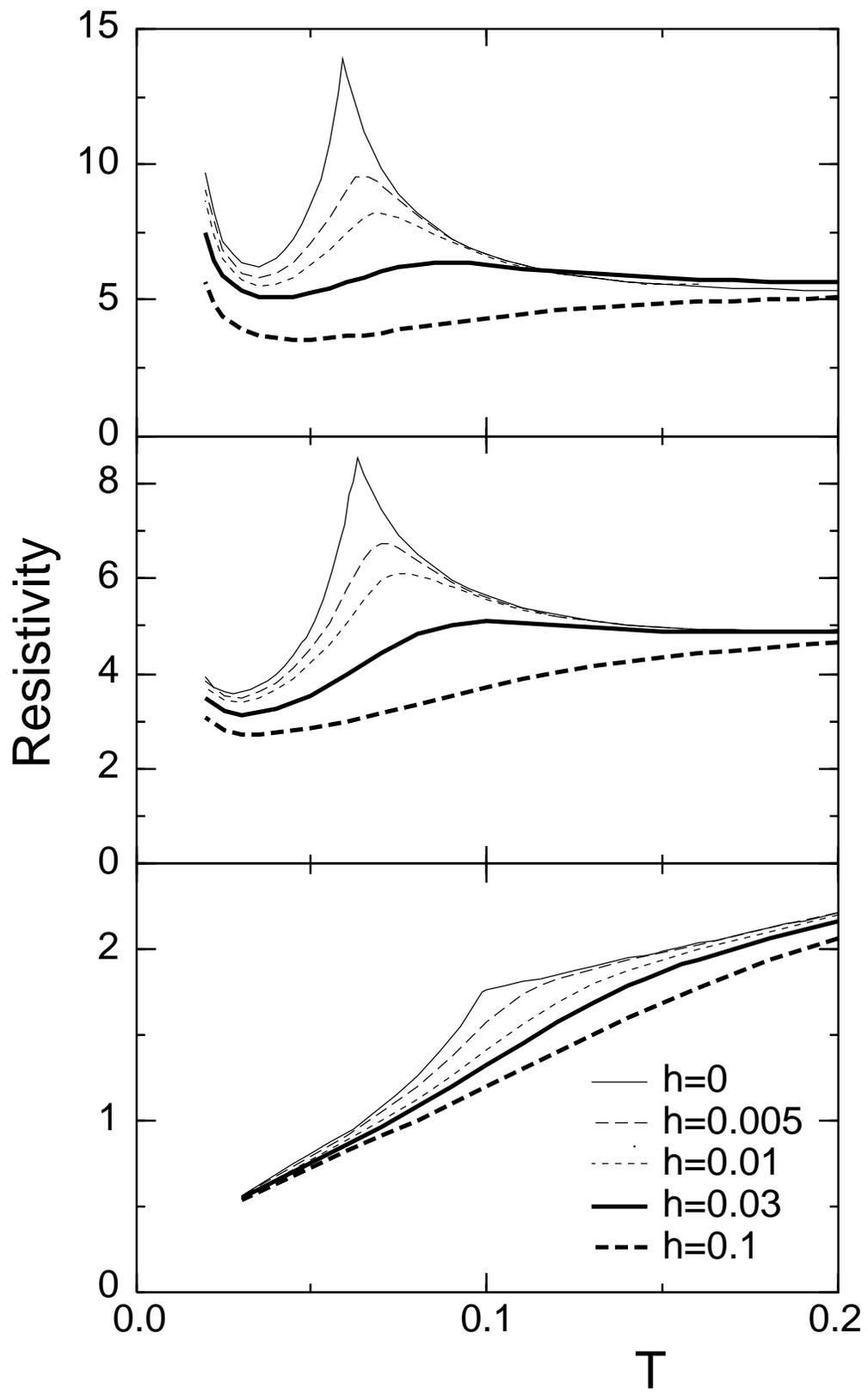
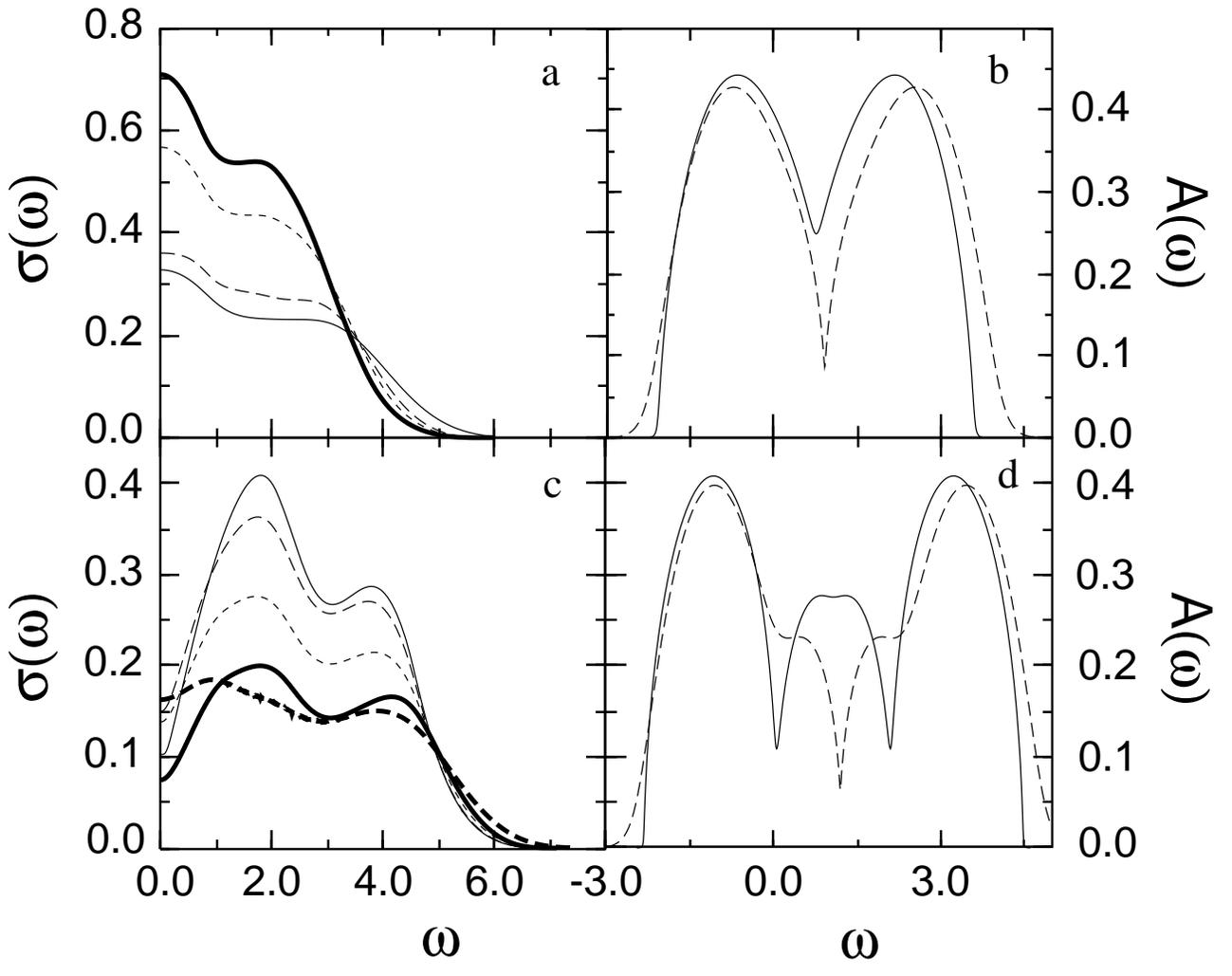
\end{document}